\documentclass{aa}

\usepackage{graphicx}
\usepackage{booktabs}
\usepackage{txfonts}
\usepackage{lipsum}
\usepackage{subcaption}         
\usepackage{lscape}             
\usepackage{placeins}           
\usepackage{amsmath}	
\usepackage{hyperref} 

\begin{document}

   \title{Where does the simplified stellar contamination model fail in exoplanet transmission spectroscopy?}


    \author{V.Y.D. Sumida\inst{1}\inst{2}
    \and R. Estrela\inst{1}
    \and M. Swain\inst{1}
    \and A. Valio\inst{2}}

    \institute{Jet Propulsion Laboratory, California Institute of Technology, Pasadena, California, USA\\
    \and Center for Radio Astronomy and Astrophysics Mackenzie (CRAAM), Mackenzie Presbyterian University, Rua da Consolac\~ao, 930, S\~ao Paulo, S\~ao Paulo, 01302-907, Brazil\\
    \email{viktor.sumida@outlook.com}}

   \date{Accepted January 5 2026}

 
  \abstract
{
Stellar photospheric heterogeneities (e.g., starspots and faculae) distort the apparent stellar spectrum during a transit and imprint wavelength-dependent biases on the measured planet–to–star radius ratio. This Transit Light Source Effect (TLSE) must be accounted for to obtain reliable atmospheric properties. A widely used approach is the Rackham–TLSE (R–TLSE) prescription, which applies a disc-averaged contamination correction based solely on the filling factor and spectral contrast. However, accurate transmission-spectroscopy interpretations require models that also account for limb darkening, the spatial distribution of active regions, and transit geometry. In this work, we incorporate these effects into a self-consistent, pixel-resolved framework, \texttt{ECLIPSE-X$\lambda$}, and first perform idealised, noise-free model–model comparisons against the R–TLSE approximation. Using three archetypal systems – the super-Earth LHS~1140\,b, the mini-Neptune K2-18\,b, and the hot Jupiter WASP-69\,b – we show that disc-averaged TLSE corrections can differ from the self-consistent model by up to $\sim$400~ppm in the optical for active hosts and non-equatorial transits, while remaining below $\sim$10~ppm at near-infrared wavelengths where limb darkening is weaker. We then apply both approaches to the JWST/NIRISS SOSS transmission spectrum of LHS~1140\,b. When limb darkening is artificially set to zero, \texttt{ECLIPSE-X$\lambda$} recovers stellar-contamination parameters that closely match the reference R–TLSE solution, confirming that both frameworks are consistent in the disc-averaged limit. With wavelength-dependent limb darkening included, however, reproducing the observed short-wavelength slope through stellar contamination alone requires very hot faculae ($\Delta T_{\mathrm{fac}}\simeq600$\,K) covering $f_{\mathrm{fac}}\simeq0.35$ of the visible hemisphere, corresponding to an equivalent circular facular region with radius $\simeq0.6\,R_\star$ (i.e., about 60\% of the stellar radius) on the stellar disc. Such an extended, unocculted active region would already be physically unlikely even for an active M dwarf. In this context, a purely stellar-contamination explanation for any residual optical slope would demand rather extreme facular populations, whereas scenarios in which a genuine atmospheric contribution helps to complement a more modest facular signal appear more physically plausible. Taken together, these results delineate the regime of validity of the R–TLSE approximation and underscore the need for geometry-aware stellar-heterogeneity models that explicitly account for limb darkening in high-precision transmission spectroscopy.}


   \maketitle

\section{Introduction} 
\label{sec:intro}

Transmission spectroscopy has become a cornerstone technique for probing exoplanetary atmospheres, enabling the detection of chemical species, clouds, and scattering phenomena \citep{Seager.and.Sasselov.2000ApJ...537..916S, Brown.et.al.2001ApJ...552..699B, Charbonneau.et.al.2002ApJ...568..377C}. However, interpretation of these observations can be made more challenging by the presence of unocculted stellar photospheric heterogeneities (i.e., starspots and faculae), which distort the host star's spectrum during transits and can mimic, or obscure, planetary atmospheric features \citep{McCullough.et.al.2014ApJ...791...55M, Herrero.et.al.2016A&A...586A.131H, Rackham.et.al.2017ApJ...834..151R, Rackham.et.al.2018ApJ...853..122R, Rackham.et.al.2019, Chachan.et.al.2019AJ....158..244C, Iyer.and.Line.2020ApJ...889...78I}. 

This source of systematic distortion, sometimes termed the Transit Light Source Effect \citep[TLSE,][]{Rackham.et.al.2018ApJ...853..122R}, can produce apparent spectral modulation that surpasses the amplitude of genuine planetary atmospheric signals, particularly in systems with active M-dwarfs and K-type stars. Although the James Webb Space Telescope (JWST) has unprecedented sensitivity to molecules such as CH$_4$, CO$_2$, and H$_2$O, wavelength-dependent flux deficits introduced by unocculted spots and faculae can mimic molecular absorption bands, mask genuine spectral features, and bias the retrieved planet-to-star radius ratio, thereby complicating the interpretation of transmission spectra.

\citet{Rackham.et.al.2018ApJ...853..122R} introduced an analytic formalism -- hereafter referred to as the Rackham-TLSE (R-TLSE) -- which has since become the default means of mitigating stellar heterogeneity in atmospheric retrievals \citep[e.g.,][]{Lim.et.al.2023ApJ...955L..22L, May.et.al.2023ApJ...959L...9M, Moran.et.al.2023ApJ...948L..11M, Cadieux.et.al.2024ApJ...970L...2C, Fournier-Tondreau.et.al.2024MNRAS.528.3354F,
Petit.dit.de.la.Roche.et.al.2024A&A...692A..83P,
Radica.et.al.2025ApJ...979L...5R, Saba.et.al.2025ApJS..276...70S, Perdelwitz.et.al.2025ApJ...980L..42P}.
Although the R-TLSE formalism has become a widely used approximation to model stellar contamination, it remains a simplified treatment of the effects of photospheric inhomogeneity. This approach describes the observed transit depth as a function of the contrast between the spectrum of the pristine photosphere ($F_{\lambda,\mathrm{phot}}$), the spectrum of heterogeneity ($F_{\lambda,\mathrm{het}}$), and the fractional coverage of the stellar disc by active regions (the filling factor). However, it does not account for the geometry of the transit chord, the stellar limb-darkening profile, and the spatial distribution of active regions across the stellar disc. As a result, atmospheric retrievals based solely on the TLSE formalism may be biased, particularly for systems with complex stellar activity that are observed with high-precision measurements. 

In this work, we present a comparative analysis between the R-TLSE approximation and our self-consistent modelling framework, \texttt{ECLIPSE-X$\lambda$}. Using three representative exoplanetary systems -- LHS\,1140\,b (a super-Earth), K2-18\,b (a mini-Neptune), and WASP-69\,b (a hot Jupiter) -- we systematically quantify the extent to which these approximations influence spectral interpretations. Our findings clearly identify the conditions under which the R-TLSE approximation becomes unreliable, emphasising the necessity for rigorous modelling of stellar activity to robustly disentangle planetary atmospheric signals from stellar contamination. Conversely, we also delineate the conditions under which the R-TLSE approximation remains valid, clarifying its domain of applicability.

The structure of this paper is as follows. In Section~\ref{sec:methods}, we describe in detail the implementation of the \texttt{ECLIPSE-X$\lambda$} model and the treatment of the Rackham–TLSE prescription. In Section~\ref{sec:results}, we first present an idealised, noise-free comparison that employs a differential error metric to quantify discrepancies between the two formalisms for the three archetypal systems aforementioned. We then apply both frameworks to the JWST/NIRISS SOSS transmission spectrum of LHS\,1140\,b to assess their performance in a realistic observational context. We also discuss the implications of these findings for exoplanet atmosphere characterisation, focussing on the regimes in which the R--TLSE correction loses validity while noting the specific conditions where it remains a useful simplification. The main conclusions are summarised in Section~\ref{sec:conclusions}. Supplementary figures and retrieval outputs for K2-18\,b and WASP-69\,b are provided in Section~\ref{sec:appendix}.

\section{Methods}
\label{sec:methods}

\subsection{The \texttt{ECLIPSE-X$\lambda$} starspot model}\label{sec:starspot.model}

The \texttt{ECLIPSE-X$\lambda$}\footnote{\url{https://github.com/ViktorSumida/ECLIPSE-Xlambda}} \citep[][v1.0.0, Zenodo, doi:\href{https://zenodo.org/records/10888850}{10.5281/zenodo.10888850}, as developed on \href{https://github.com/ViktorSumida/ECLIPSE-Xlambda}{GitHub}]{ECLIPSE-Xlambda} is an advanced transit modelling tool built upon the foundation of the \texttt{ECLIPSE}\footnote{\url{https://github.com/Transit-Model-CRAAM/}}  model, originally proposed by \cite{Silva.2003}.
The code expanded the functionality of \texttt{ECLIPSE} allowing for simulation of transit light curves across a wide range of wavelengths.
The original \texttt{ECLIPSE} code was primarily designed to model active-region occultations using a 2D image of a synthetic star with spots on the surface of the limb-darkened stellar disc. By analysing the photometric variations in the observed light curves during exoplanet transits, the physical properties of these systems can be inferred \citep[e.g.,][]{Silva-Valio.et.al.2010, Valio.et.al.2017, Zaleski.et.al.2019, Zaleski.et.al.2020, zaleski.et.al.2022, Netto.and.Valio.2020, Araujo.and.Valio.2021, valio2022, Valio2024}. 

To produce synthetic light curves, \texttt{ECLIPSE-X$\lambda$} requires a 3D matrix to store stellar disc data at given wavelengths.
This matrix encapsulates the spatial information of the star's surface, enabling the code to simulate the effect of the transit and accurately compute the resulting flux variations. \texttt{ECLIPSE-X$\lambda$}  generates multiwavelength synthetic light curves that closely resemble the observed data, and allows calculation of the transit depth which in turn enables the analysis of the atmospheric properties of the exoplanetary system under investigation.
The code is primarily written in Python although functions in \texttt{C} were implemented to enhance computational efficiency.
Next, we detail the relevant input parameters of \texttt{ECLIPSE-X$\lambda$}.  Additional input parameters and their details can be found in \cite{Silva.2003}.

Each starspot or facula is modelled as a circular disc and foreshortening of the spots and faculae is taken into account when they are close to the stellar limb. The spot size parameter is measured as a fraction of the stellar radius ($R_\mathrm{s}$), and its position, namely latitude and longitude, is given with respect to the centre of the stellar disc. The relative area of the stellar disc covered by starspots is the filling factor (hereafter \textit{f\,f}), defined as
\begin{equation}
    f\!f = \frac{A_\mathrm{spot/fac}}{A_\mathrm{s}} = \left( \frac{R_\mathrm{spot/fac}}{R_\mathrm{s}} \right) ^ 2 \; ,
\end{equation}
where $A_\mathrm{s}$ and $R_\mathrm{s}$ are the area and radius of the stellar disc, and $A_\mathrm{spot/fac}$ and $R_\mathrm{spot/fac}$ are area and radius of the starspot or facula, respectively.

\subsection{Constructing wavelength-dependent contrasts}
\label{sec:continuum_line}

In both the \texttt{ECLIPSE-X$\lambda$} and R--TLSE (see Section~\ref{sec:R-TSLE}) frameworks, stellar contamination can be expressed in terms of the wavelength-dependent contrast between the heterogeneous region and the pristine photosphere,
\begin{equation}
    r_\lambda \equiv \frac{F_{\lambda,\mathrm{het}}}{F_{\lambda,\mathrm{phot}}} \; ,
    \label{eq:r_lambda}
\end{equation}
where $F_{\lambda,\mathrm{het}}$ and $F_{\lambda,\mathrm{phot}}$ are the disc-integrated fluxes from the heterogeneous region and the immaculate photosphere, respectively. 
When $r_\lambda$ is computed directly from high-resolution stellar-atmosphere spectra, for instance by interpolating PHOENIX specific intensities \citep{Hauschildt.et.al.1997ApJ...483..390H,Hauschildt.et.al.1999JCoAM.109...41H,Husser.et.al.2013} with LDTK \citep{Parviainen.and.Aigrain.2015MNRAS.453.3821P}, it exhibits sharp, rapidly varying structure associated with atomic and molecular lines.
Within a pixelated transit model that already includes wavelength-dependent limb darkening, feeding these line-dominated contrasts directly into the calculation leads to a strong and largely artificial amplification of individual spectral lines: the effective limb-darkening coefficients track these narrow features (see \autoref{fig:LDCs}) and imprint them onto the transit depth in a way that is not present in the original one-dimensional atmosphere models.

To prevent this line-amplification problem, in this work we decouple the smooth continuum behaviour of the stellar intensity from the detailed line structure. For the computation of limb-darkening coefficients, we do not use monochromatic \texttt{PHOENIX} intensities, as this would propagate line structure directly into the limb-darkening law. Instead, we adopt a plane-parallel grey-atmosphere approximation in radiative equilibrium to obtain a spectrally smooth centre-to-limb intensity profile. For each wavelength bin, we compute the emergent intensity $I_\lambda(\mu)$ as a function of $\mu=\cos\theta$, normalise by $I_\lambda(1)$, and fit the four-parameter limb-darkening law of \citet{Claret.2000}. The best-fitting coefficients are obtained by minimising the squared residuals between the grey-atmosphere intensity profile and the parametric law using a Levenberg--Marquardt non-linear least-squares algorithm, following the same optimisation strategy adopted by \cite{Claret.et.al.2025A&A...699A..97C} for \texttt{PHOENIX}-based JWST passbands. This yields a set of limb-darkening coefficients that varies smoothly with wavelength and is free from narrow spectral-line signatures. The same grey-atmosphere limb-darkening law is used for both the immaculate photosphere and the active regions, while their relative brightness is set by the black-body contrast:
\begin{equation}
    \frac{I_{\mathrm{spot/fac}}}{I_{\mathrm{star}}} =
    \frac{\exp\,\bigl[h c / (\lambda K_\mathrm{B} T_{\mathrm{star}})\bigr] - 1}
         {\exp\,\bigl[h c / (\lambda K_\mathrm{B} T_{\mathrm{spot/fac}})\bigr] - 1} \; ,
    \label{eq:bb_contrast}
\end{equation}
where $h$ is the Planck constant, $K_\mathrm{B}$ is the Boltzmann constant, $T_{\mathrm{star}}$ is the stellar effective temperature, and $T_{\mathrm{spot/fac}}$ is the temperature of the starspot or facula.

At the same time, we seek to preserve the physical fidelity of the spectral-line pattern. To accomplish this, we decompose the contrast into a smoothly varying continuum component and a high-frequency residual. This separation is most conveniently performed in logarithmic space,
\begin{equation}
    \log r_\lambda \;=\; \log r_{\lambda,\mathrm{LP}} \;+\;
    \bigl[\log r_\lambda - \log r_{\lambda,\mathrm{LP}}\bigr] \; ,
    \label{eq:logr_decomposition}
\end{equation}
where $\log r_{\lambda,\mathrm{LP}}$ denotes a low-pass filtered version of the contrast (i.e., the smooth envelope or continuum), and the bracketed term contains the high-frequency residual 
$\Delta \log r_\lambda = \log r_\lambda - \log r_{\lambda,\mathrm{LP}}$,
which encodes only the contribution from spectral lines. In practice, we compute $r_\lambda$ from \texttt{PHOENIX} stellar spectra for the photosphere and for the active regions, apply a low-pass filter in wavelength to obtain $\log r_{\lambda,\mathrm{LP}}$, and identify the residual $\Delta \log r_\lambda$ as the purely line-induced component.

For the self-consistent \texttt{ECLIPSE-X$\lambda$} simulations, the smooth component $r_{\lambda,\mathrm{LP}}$ is then replaced by the black-body contrast implied by \autoref{eq:bb_contrast}, combined with the aforementioned grey-atmosphere limb-darkening description, while the residual $\Delta \log r_\lambda$ is applied as a multiplicative, wavelength-dependent correction at the level of the disc-integrated fluxes. In this way, the contaminated signal predicted by \texttt{ECLIPSE-X$\lambda$} is governed by a spectrally smooth continuum and physically motivated limb darkening, while the fine spectral structure of realistic stellar spectra is still consistently incorporated into the model.

\vspace{.7cm}

\subsection{R-TLSE model comparison}
\label{sec:R-TSLE}

According to \cite{Rackham.et.al.2018ApJ...853..122R}, the parameter $\epsilon_{\lambda,\text{het}}$ represents the correction factor introduced by the TLSE. It quantifies the discrepancy between the observed transit depth and the intrinsic transit depth expected in the absence of stellar heterogeneities at a given wavelength $\lambda$. The factor $\epsilon_{\lambda}$ is defined as the ratio of the observed transit depth to the intrinsic transit depth, accounting for the presence of unocculted stellar spots and faculae. Mathematically, it is expressed as:

\begin{equation}
\epsilon_{\lambda,\text{het}} = \frac{1}{1 - f_{\text{het} \left( 1 - \frac{F_{\lambda,\text{het}}}{F_{\lambda,\text{phot}}} \right)}} \; .
\label{eq:epsilon_TSLE}
\end{equation}

If $\epsilon_{\lambda} > 1$, the measured transit depth is overestimated, indicating an excess of dimmer regions on the stellar surface (e.g., unocculted starspots). Conversely, if $\epsilon_{\lambda} < 1$, the transit depth is underestimated, suggesting an excess of brighter regions, such as faculae.

In this study, we compare the simple R-TLSE model proposed by \cite{Rackham.et.al.2019} to our TLSE model, which incorporates additional physical effects.
The comparison in this study will be carried out in a straightforward and systematic manner. First, we simulate two scenarios: a pristine, unspotted star (serving as a baseline) and a star with heterogeneous surface features (incorporating stellar contamination). For the TSLE, we will use the unspotted star as a reference and infer the stellar contamination through the contamination factor $\epsilon_{\lambda,\text{het}}$, allowing for a direct evaluation against the theoretical predictions of the R-TLSE model. To quantify discrepancies between the frameworks, we introduce the error:
\begin{equation}
    \text{Error} = D_{\lambda, \text{ECLIPSE}} - D_{\lambda, \text{R-TSLE}} \;,
\label{eq:err}
\end{equation}
where \( D_{\lambda, \text{ECLIPSE}} \) is the transit depth obtained from our simulations, and \( D_{\lambda, \text{R-TSLE}} \) is the transit depth computed using \autoref{eq:epsilon_TSLE} of R-TSLE \citep{Rackham.et.al.2018ApJ...853..122R}. This metric isolates systematic biases introduced by the R-TLSE’s simplified treatment of limb darkening and transit geometry.
Positive values of \(\text{Error}\) indicate that R--TLSE predicts a shallower transit than ECLIPSE-X$\lambda$ (i.e., it overcorrects for stellar contamination), whereas negative values indicate that R--TLSE predicts a deeper transit and thus undercorrects for stellar contamination.

To guide the interpretation of the forthcoming detailed comparison in Section~\ref{sec:results}, \autoref{tab:model_comparison} provides a concise overview of the conceptual differences between the R–TLSE and ECLIPSE-X$\lambda$ frameworks, particularly regarding limb darkening, transit geometry, and the treatment of stellar surface heterogeneities.

\begin{table*}
\caption{
Qualitative comparison between the R--TLSE formalism and the ECLIPSE-X$\lambda$.
}
\label{tab:model_comparison}
\centering
\renewcommand{\arraystretch}{1.2}
\begin{tabular}{l p{0.36\textwidth} p{0.36\textwidth}}
\hline\hline 
\addlinespace[0.2em]
Conceptual element &
R--TLSE &
ECLIPSE-X$\lambda$ \\[0.2em]
\hline
\addlinespace[0.2em]
Model formulation &
Disc-averaged, one-dimensional correction factor
$\epsilon_{\lambda}$ applied to the intrinsic transit depth. &
Pixel-resolved 2D stellar disc; transit depth obtained from explicit
occultation of a limb-darkened star with embedded active regions. \\[0.4em]

Limb darkening &
No explicit limb darkening; the stellar disc is treated as
uniformly bright. &
Includes wavelength-dependent limb darkening via a
four-parameter law. \\[0.4em]

Transit geometry &
Absent from the R--TLSE contamination factor; the same
$\epsilon_{\lambda}$ is used for any impact parameter. &
Full transit geometry (impact parameter)
is treated self-consistently in the pixel-based light-curve
calculation. \\[0.4em]

\parbox[t]{3.5cm}{Location of active regions on the stellar disc} &
Only the total filling factor matters; the spatial distribution
of spots and faculae across the disc is not resolved. &
Active regions are placed at specific latitudes and longitudes;
their projected position on the disc, together with limb darkening
and foreshortening effects, modulates the amplitude and wavelength
dependence of the contamination signal. \\[0.4em]

Spotless configuration &
In the absence of active regions the contamination factor is
unity and the transmission spectrum is strictly flat. &
Even for a spotless star the transmission spectrum can show
a mild positive or negative slope in $R_{\rm p}/R_\star(\lambda)$,
driven solely by limb darkening and transit geometry. \\[0.4em]

Starspots &
Unocculted cool spots always imprint a net increase in
transit depth towards shorter wavelengths, i.e. a positive
optical slope. &
Cool spots tend to drive a positive slope, but the net trend can
be weakened, cancelled, or even reversed depending on the
impact parameter. \\[0.4em]

Faculae &
Unocculted faculae always reduce the transit depth towards
shorter wavelengths, producing a negative optical slope. &
Faculae tend to generate a negative slope, but the final
wavelength dependence results from the competition between
facular contrast and the geometry–induced trend, and can be
substantially weaker or of opposite sign. \\[0.2em]
\hline
\end{tabular}
\end{table*}

\section{Results and discussion}
\label{sec:results}

This section combines an idealised, purely model–model comparison (Section~\ref{sec:idealised_model_comparison}) with an application to observational data (Section~\ref{sec:observational_LHS1140b}). In Section~\ref{sec:idealised_model_comparison} (Cases~1–3), we compute noise-free transmission spectra with \texttt{ECLIPSE-X$\lambda$} and with the R--TLSE approximation using the same black-body description for the photosphere and active regions, without including \texttt{PHOENIX} stellar-atmosphere spectra or any instrument response. This common black-body prescription avoids mixing a one-dimensional \texttt{PHOENIX} atmosphere with the pixelised stellar disc in \texttt{ECLIPSE-X$\lambda$} and ensures that the discrepancies we report are driven solely by the treatment of stellar heterogeneity and transit geometry, rather than by differences in radiative-transfer inputs or spectral binning. In Section~\ref{sec:observational_LHS1140b}, when we apply both approaches to the JWST/NIRISS SOSS observations of LHS~1140\,b, we instead adopt \texttt{PHOENIX} model atmospheres for both the photosphere and the active regions and fold the models through the NIRISS throughput and spectral resolution, so that the comparison reflects a realistic observational context and the impact of these systematic differences can be assessed relative to the actual measurement uncertainties.

\subsection{Idealised model–model comparison} \label{sec:idealised_model_comparison}

To assess how our model deviates from the R-TLSE in its predictions, we performed simulations under the following scenarios:

\begin{itemize}
\item Case~1 -- Fixed filling factor, variable temperature contrast: The filling factor was set to 0.15 for both starspots and faculae, while the temperature contrast $\Delta T$ was varied from $-300$\,K to $+300$\,K. This disentangles spectral biases driven purely by temperature differences.
The adopted upper limit for the filling factor ($f_{\rm het} = 0.15$) is motivated by recent atmospheric retrievals based on JWST transmission spectra, which inferred comparable starspot and faculae coverages in M-dwarf systems. Specifically, \cite{May.et.al.2023ApJ...959L...9M} and \cite{Cadieux.et.al.2024ApJ...970L...2C} reported coverages of up to 40\% for GJ\,1132\,b and 30\% for LHS\,1140\,b, based on NIRSpec and NIRISS data, respectively.

\item Case~2 -- Fixed temperature contrast, variable filling factor: A constant temperature contrast of $|\Delta T| = 300$\,K was applied, while the filling factor was varied from 0.01 to 0.15. This isolates the effect of the spatial coverage of active regions.
We constrained the temperature contrast between active regions and the photosphere to $\pm300$\,K, consistent with values inferred from those retrievals \citep{May.et.al.2023ApJ...959L...9M, Cadieux.et.al.2024ApJ...970L...2C} and from solar analogue studies \citep[e.g.,][]{Meunier.et.al.2010A&A...512A..39M}.

\item Case~3 -- Variable position of active regions: We fixed both the temperature contrast and filling factor, but systematically varied the projected position of starspots and faculae on the stellar disc by changing their latitude and longitude. This evaluates the geometric dependence of stellar contamination and its impact on spectral bias.
\end{itemize}

\begin{figure}
    \centering
    \includegraphics[width=0.99\columnwidth]{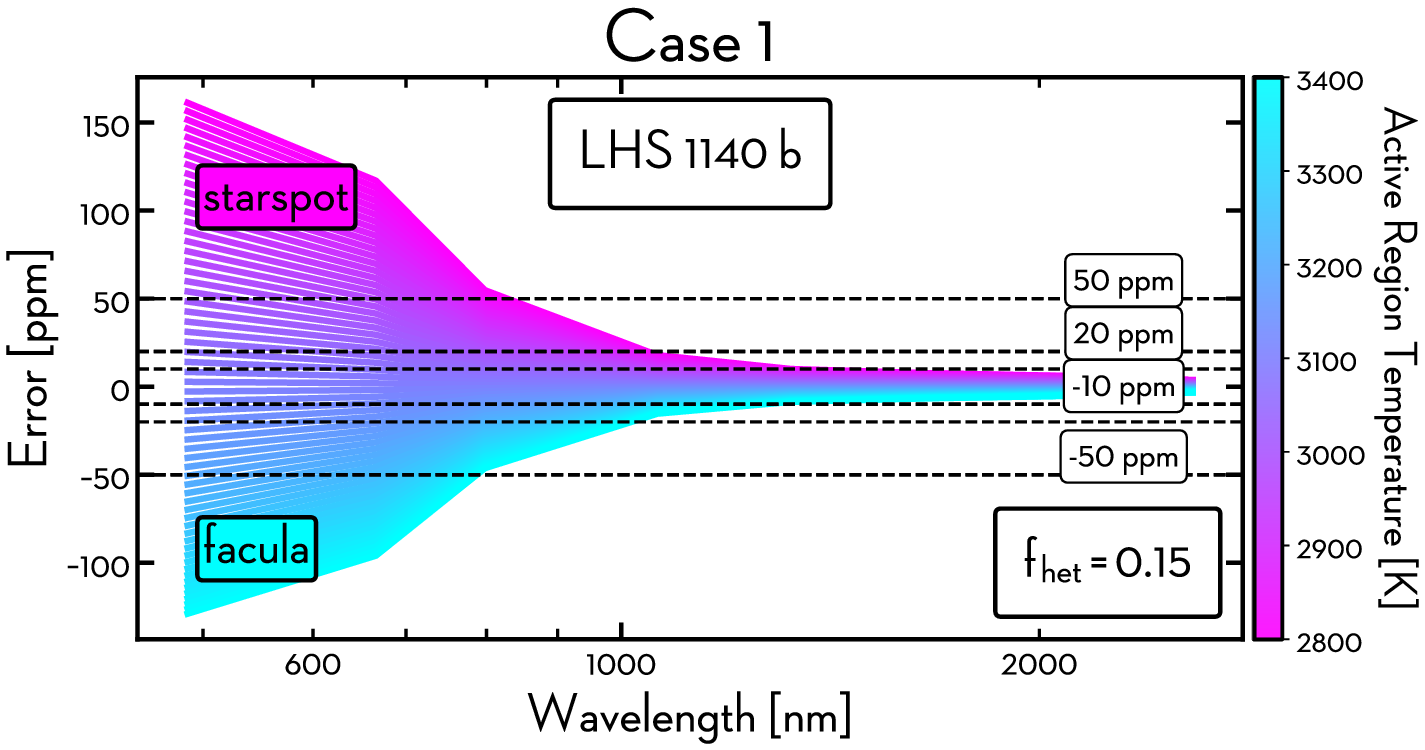}
    \caption{Wavelength-dependent errors between the \texttt{ECLIPSE-X$\lambda$} and the R-TLSE (\autoref{eq:err}) for LHS\,1140\,b, computed using a fixed filling factor ($f_{\rm het} = 0.15$) and a temperature contrast ($\Delta T$) varying from $-300$\,K (spots) to $+300$\,K (faculae). The largest discrepancies arise in the optical range (500--800\,nm), where steep limb-darkening gradients drive optical biases. The dashed lines mark reference levels at -50, -20, -10, 10, 20, and 50\,ppm, serving as a visual aid to highlight variations in the data.
    By construction, positive values of the error indicate that R--TLSE predicts a shallower transit depth than ECLIPSE-X$\lambda$ (overcorrecting stellar contamination), while negative values correspond to a deeper R--TLSE transit (undercorrecting contamination).} \label{fig:LHS1140b_err_lambda_ff_0p15.eps}
    \end{figure}

\begin{figure}
    \centering
    \includegraphics[width=0.99\columnwidth]{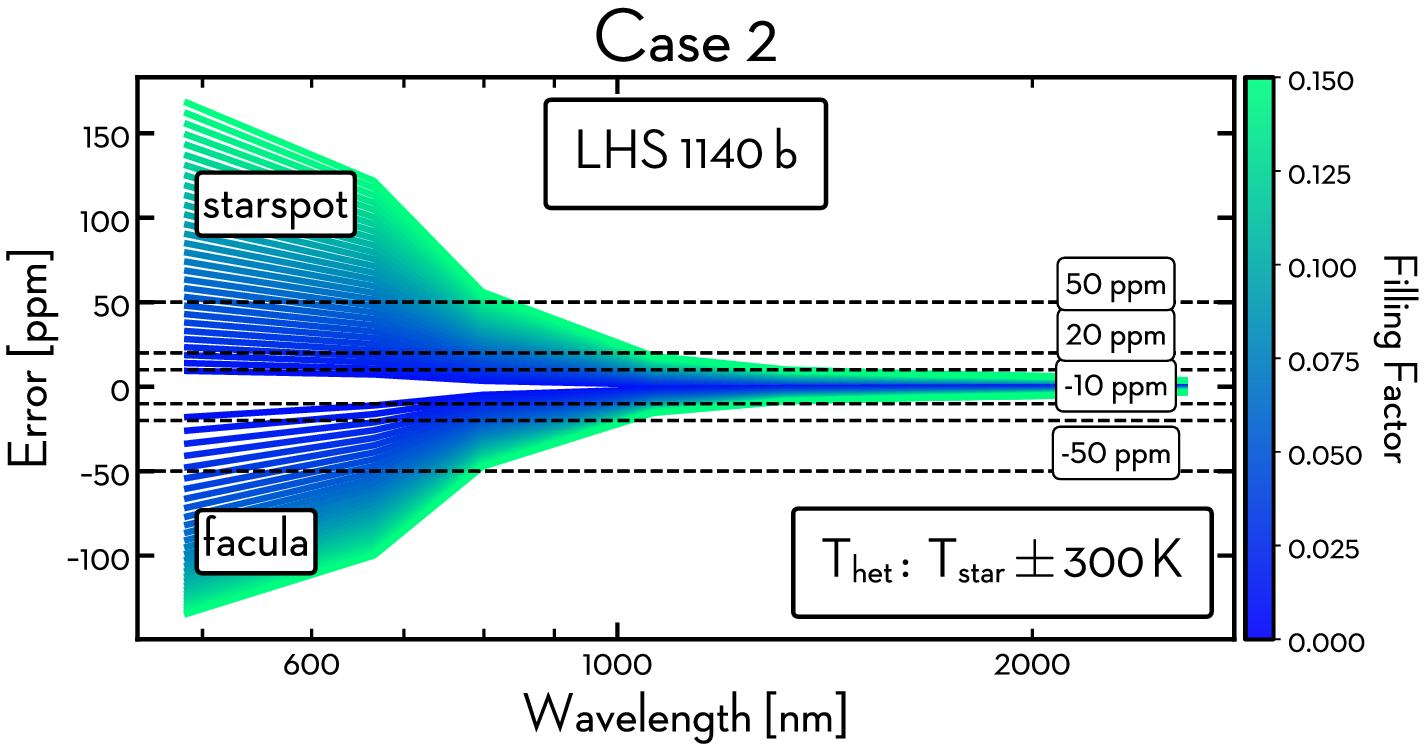}
    \caption{Discrepancies, defined by \autoref{eq:err}, as a function of fixed temperature contrast ($T_\mathrm{star}\pm300$) and vary the filling factor ($f_\mathrm{het}$) from 0.01 to 0.15. Positive errors (R-TLSE overestimation) dominate for cool spots, while negative errors (R-TLSE underestimation) arise for hot faculae. Extreme cases (e.g., $f_{\rm het} = 0.15$, $\Delta T = \pm 300$ K) show maximal divergence. The dashed lines delineate boundaries at -50, -20, -10, 10, 20, and 50\,ppm.}
    \label{fig:LHS1140b_err_lambda_T_300}
\end{figure}

The results for Case~1 and 2, for the planet LHS\,1140\,b, are presented in Figures~\ref{fig:LHS1140b_err_lambda_ff_0p15.eps} and~\ref{fig:LHS1140b_err_lambda_T_300}, respectively.
Parallel analyses for K2-18\,b and WASP-69\,b are provided in Figure~\ref{fig:K2-18b_WASP-17b_err_lambda}.

The pattern observed in the curves in Figures \ref{fig:LHS1140b_err_lambda_ff_0p15.eps} and \ref{fig:LHS1140b_err_lambda_T_300} -- and Figure~\ref{fig:K2-18b_WASP-17b_err_lambda} in \autoref{sec:appendix} -- aligns with expectations: the largest wavelength-dependent errors between the R-TLSE approximation and the \texttt{ECLIPSE-X$\lambda$} model occur at wavelengths where limb darkening induces the most pronounced brightness gradients across the stellar disc, particularly in the optical regime \citep[see][]{Alexoudi.et.al.2020A&A...640A.134A}. 
For LHS\,1140\,b, our simulations reveal significant discrepancies, especially at optical wavelengths. At wavelengths shorter than 800\,nm, systematic biases exceed 50\,ppm, reaching maxima of approximately 170\,ppm for unocculted starspots and 140\,ppm for faculae near 500\,nm. 

These offsets arise from the R-TLSE’s neglect of wavelength-dependent limb-darkening contrasts and the geometric modulation of stellar contamination. 
At longer wavelengths (1000–2000 nm), the error diminishes to below 10 ppm, as limb-darkening effects weaken and stellar spectral energy distributions become smoother. Adopting the system parameters of \cite{Cadieux.et.al.2024ApJ...970L...2C}, a single atmospheric scale height of an N$_2$-dominated atmosphere on LHS~1140~b corresponds to only a few ppm (on the order of 4--5\,ppm) in transit depth; the 50--170\,ppm optical discrepancies between ECLIPSE-X$\lambda$ and the R--TLSE approximation therefore amount to on the order of 10--40 atmospheric scale heights. This wavelength-dependent bias, therefore, has direct implications for atmospheric retrievals and for the interpretation of slope-like features in terms of Rayleigh scattering or aerosols. Previous TLSE-based retrieval studies have already shown that specific stellar modelling assumptions and particular treatments of stellar contamination can imprint or distort such apparent slopes, biasing the inferred atmospheric properties \citep[see, for example,][]{Zhang.et.al.2018AJ....156..178Z, Iyer.and.Line.2020ApJ...889...78I, Rackham.and.deWit.2024AJ....168...82R}.

In their study, \citet{Cadieux.et.al.2024ApJ...970L...2C} report a marginal detection of N$_2$ in the atmosphere of the temperate super-Earth LHS\,1140\,b at a significance level of 2.3$\sigma$. This result is primarily driven by a modest short-wavelength spectral slope (500–800\,nm), which they interpret as Rayleigh scattering from a high-MMW atmosphere dominated by N$_2$. 
Although the observed transmission spectrum remains largely flat at infrared wavelengths ($>$1\,$\mu$m), it exhibits a subtle blueward slope in the visible regime -- precisely where the R-TLSE approximation has the greatest potential for errors.
The same analysis also reveals a statistically significant detection (5.8$\sigma$) of unocculted stellar faculae, covering approximately 20\% of the stellar disc. This contamination is modelled using the R-TLSE formalism. However, despite the inclusion of this correction, the fit to the data exhibits a relatively high reduced chi-squared, hinting that the R-TLSE may not fully capture the spectral distortions induced by photospheric heterogeneities.

Given the significant discrepancies between the R-TLSE approximation and the \texttt{ECLIPSE-X$\lambda$} model, interpretations of an N$_2$-rich atmosphere based solely on this simplified treatment of the R-TLSE should be approached with caution. Notably, the spectral region supporting the N$_2$ signal coincides with wavelengths where inaccuracies inherent in the R-TLSE are expected to peak -- precisely where the observational data also show their largest uncertainties and dispersion. A more comprehensive examination of the fit and its limitations is presented in Section~\ref{sec:observational_LHS1140b}.

For the sub-Neptune K2-18\,b (see Figure~\ref{fig:K2-18b_WASP-17b_err_lambda}, first and second rows), our simulations show that in the optical range (500--600\,nm), the R-TLSE approximation and our self-consistent model can differ by about 80--100\,ppm. Such offsets are comparable to the amplitude of key atmospheric signals in sub-Neptune spectra. Beyond 1.5\,$\mu$m, however, these discrepancies drop below 10\,ppm, indicating that the R-TLSE remains reasonably accurate in the near-infrared, consistent with previous investigations that robustly detected methane (CH$_4$) via R-TLSE-based equation \citep[e.g.,][]{Madhusudhan.et.al.2023ApJ...956L..13M, Schmidt.et.al.2025arXiv250118477S}.
\cite{Schmidt.et.al.2025arXiv250118477S} incorporated stellar contamination models ($f_{\mathrm{het}} = 5^{+3}_{-2}\,\%$ at $\Delta T \approx 500 \pm 350$\,K cooler than the photosphere), showing that while these corrections minimally affect CH$_4$ inferences, they are crucial to avoid false positives in trace species such as CO$_2$. 

In contrast, \cite{Madhusudhan.et.al.2023ApJ...956L..13M} found no detectable stellar activity in M dwarfs hosting Hycean candidates. 
However, the apparently quiescent behaviour of M-type hosts such as K2-18 may be partly driven by orbital geometry: for transits probing mid-latitude chords ($\sim40^\circ$), our \texttt{ECLIPSE-X$\lambda$} simulations show that a spotless photosphere already yields nearly wavelength-independent transit depths, effectively removing most of the geometrical slope. In this “flat-geometry” regime, even modest chromatic stellar contamination or weak atmospheric gradients can produce shallow slopes that are comparable to the current observational uncertainties, making it difficult to distinguish genuinely high-MMW, cloud-dominated atmospheres from cases where residual stellar heterogeneity and geometry conspire to flatten the observed spectrum. This underscores the need for stellar-heterogeneity models that are explicitly consistent with the system’s orbital geometry. A more detailed discussion of the impact of limb darkening and transit geometry on transmission spectra is provided in \cite{Alexoudi.et.al.2020A&A...640A.134A} and in Section~\ref{sec:observational_LHS1140b}.

In the case of the hot Jupiter WASP-69\,b (see Figure~\ref{fig:K2-18b_WASP-17b_err_lambda}, third and fourth rows), the differences between the R-TLSE and the \texttt{ECLIPSE-X$\lambda$} model are substantially greater than those observed for LHS\,1140\,b and K2-18\,b. Around 500\,nm, these discrepancies can reach up to $\sim$400\,ppm. Moreover, up to roughly 1500\,nm, our simulations indicate that the errors remain above 50\,ppm, eventually dropping to 50--10\,ppm in the remainder of the spectrum.

Such large discrepancies become even more significant in light of the recent VLT/FORS2 spectrophotometry of WASP-69\,b by \citet{Petit.dit.de.la.Roche.et.al.2024A&A...692A..83P}, covering 340–1100\,nm. Their retrievals, which applied the R-TLSE correction equation to account for stellar contamination, inferred an extensive facula coverage of $\sim$30\% with a temperature excess of $\Delta T \approx 230\pm70$\,K. For a K-type star, such a large facular filling factor is already somewhat extreme and is likely highly biased, given the simplified assumptions inherent to the correction method. Moreover, the planet's mid-latitude impact parameter ($b\approx0.66$) implies that the transit chord probes intermediate stellar latitudes, close to the 40--45$^\circ$ regime where our \texttt{ECLIPSE-X$\lambda$} simulations predict nearly wavelength-independent transit depths for a spotless photosphere. In this resulting “flat-geometry” configuration, most of the purely geometrical slope is removed, so the negative optical slope seen in the FORS2 transmission spectrum is expected to arise primarily from the combined effects of stellar heterogeneity and atmospheric gradients, rather than from geometry alone.

Additionally, a large facula filling factor naturally induces a negative spectral slope in the optical, as illustrated in Figure~3 of \citet{Petit.dit.de.la.Roche.et.al.2024A&A...692A..83P}. In our self-consistent \texttt{ECLIPSE-X$\lambda$} calculations, the disagreement with the R-TLSE-based predictions grows systematically with increasing facula coverage: for high filling factors, the R-TLSE prescription produces much stronger chromatic stellar contamination than the pixel-resolved model. This behaviour suggests that, when applied in the mid-latitude geometric regime of WASP-69\,b, the R-TLSE approach can substantially overpredict the level of stellar contamination and thus overestimate the facula filling factor.

Prior atmospheric analyses based on HST/STIS and WFC3 data suggested strong Rayleigh scattering slopes, which were interpreted as evidence for high-altitude aerosols \citep[e.g.,][]{Estrela.et.al.2021AJ....162...91E}. However, the steep slope reported in that study might instead result from unocculted starspots or faculae, since our contamination-free simulations yield a flat spectral baseline for WASP-69\,b. Therefore, it remains uncertain whether these optical slopes arise from genuine atmospheric properties or are artifacts of stellar heterogeneity.

Meanwhile, \citet{Murgas.et.al.2020A&A...641A.158M} reported an even steeper optical slope with GTC/OSIRIS, consistent with Rayleigh scattering, which they modelled with $f_{\text{spot}} = 0.55$ and $f_{\text{fac}} = 0.15$ ($\Delta T_{\text{spot}} = -122$\,K, $\Delta T_{\text{fac}} = +72$\,K). Such extreme filling factors almost certainly do not trace stellar heterogeneity alone, but rather reflect a degeneracy in which part of the atmospheric signal is absorbed into the stellar-contamination term by the R-TLSE formalism. For instance, a spot filling factor of $f_{\text{spot}} = 0.55$ corresponds to an equivalent circular heterogeneous region with radius $R_{\text{het}} \simeq 0.75\,R_\star$; when the facular contribution is included, the total coverage ($f_{\text{het}} = f_{\text{spot}} + f_{\text{fac}} \simeq 0.70$) implies $R_{\text{het}} \simeq 0.83\,R_\star$. In our 2D stellar model, such a vast coherent active region on the visible hemisphere would be difficult to reconcile with the mid-latitude transit chord of WASP-69\,b without the planet occulting a substantial fraction of these inhomogeneities. This strongly suggests that the steep OSIRIS slope is more naturally explained by a combination of genuine atmospheric opacity and more moderate stellar heterogeneity than by the extreme active-region coverages implied by the disc-averaged R-TLSE prescription.

Importantly, recent JWST eclipse observations of WASP-69\,b also support the presence of high-altitude aerosols on the dayside hemisphere \citep{Schlawin.et.al.2024AJ....168..104S}. Although those emission spectra are unaffected by stellar heterogeneity, the fact that cloud-free models fail to reproduce the observed flux reinforces the interpretation that aerosols, rather than stellar contamination, likely shape the near-infrared features. This highlights the need for transmission analyses to account for both stellar and planetary contributions explicitly, as done in \texttt{ECLIPSE-X$\lambda$}, which treats stellar contamination independently of atmospheric retrievals but allows for their disentanglement across wavelengths.

\begin{figure*}
    \centering
    \includegraphics[width=0.32\textwidth]{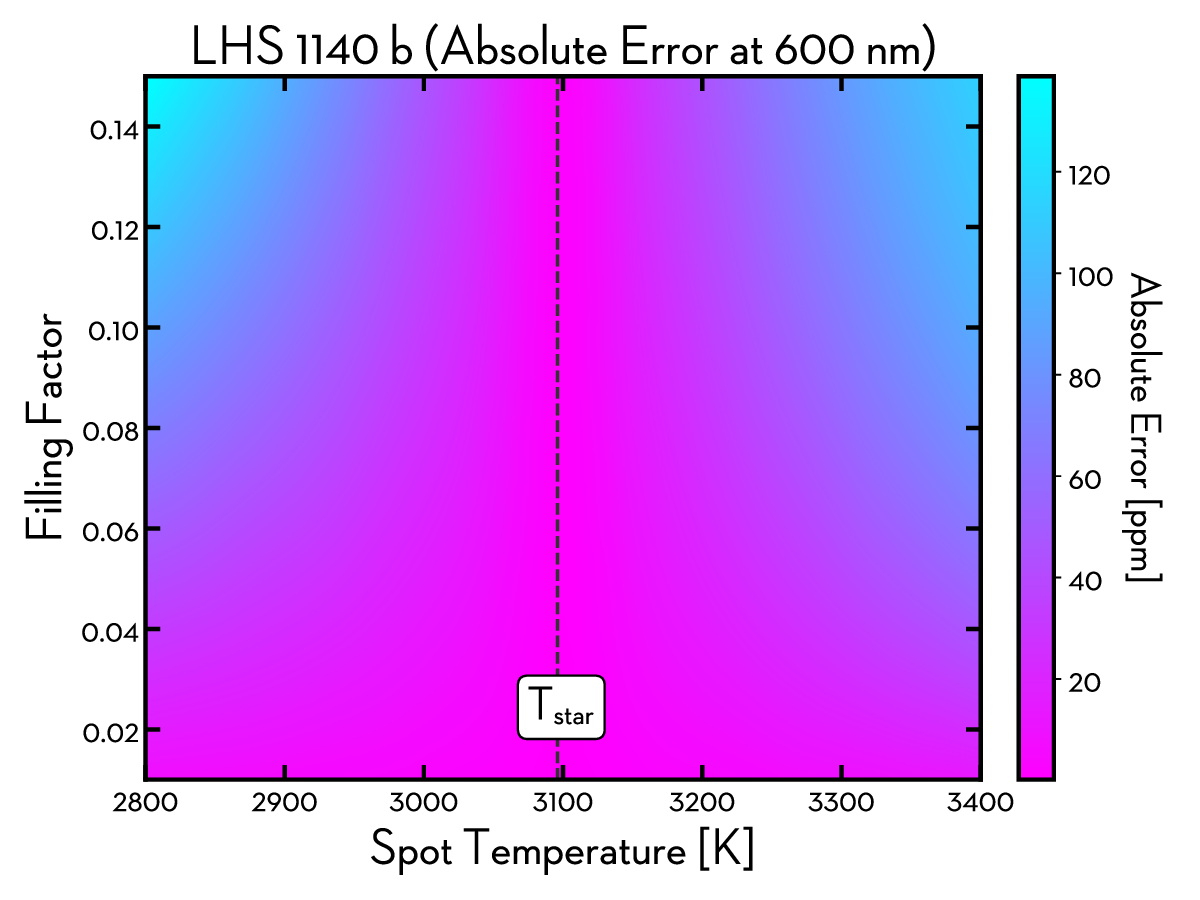}
    \includegraphics[width=0.32\textwidth]{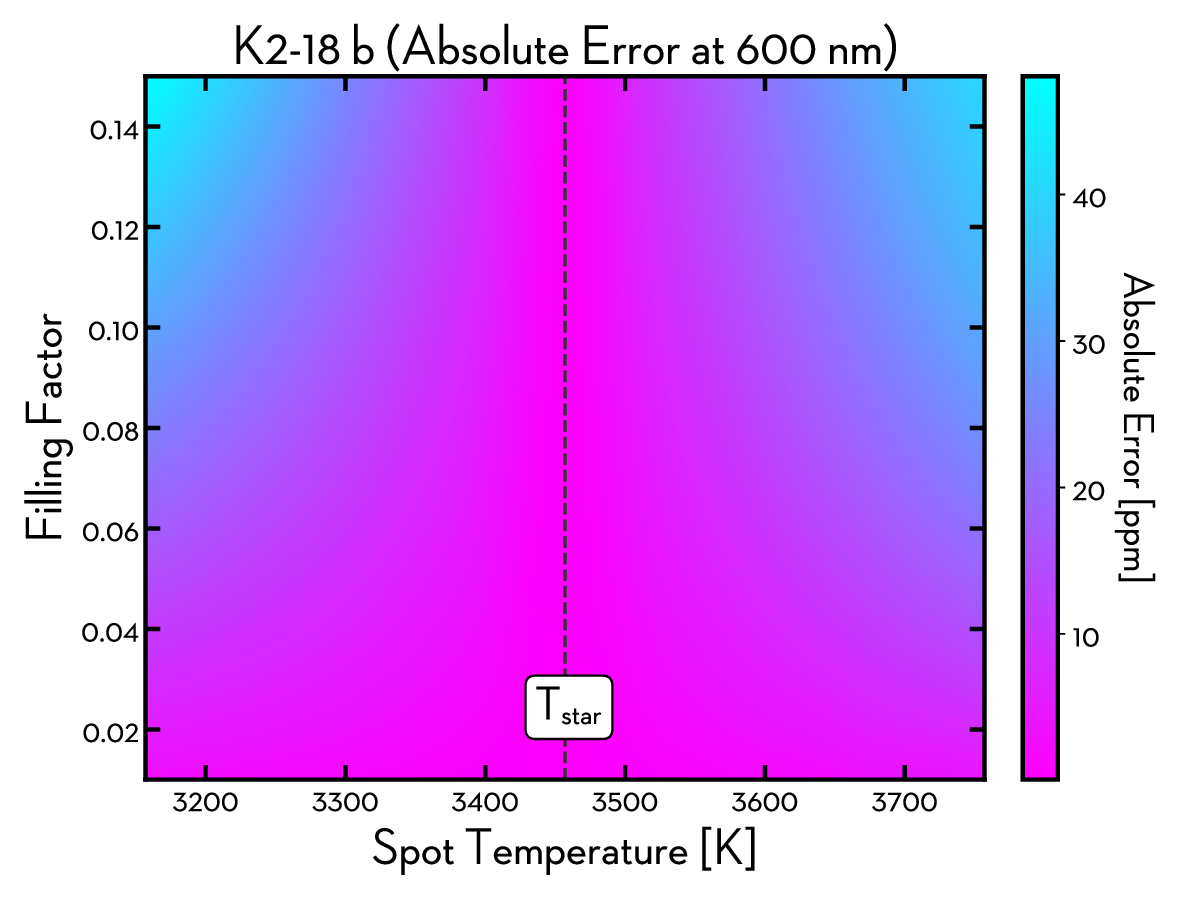}
    \includegraphics[width=0.32\textwidth]{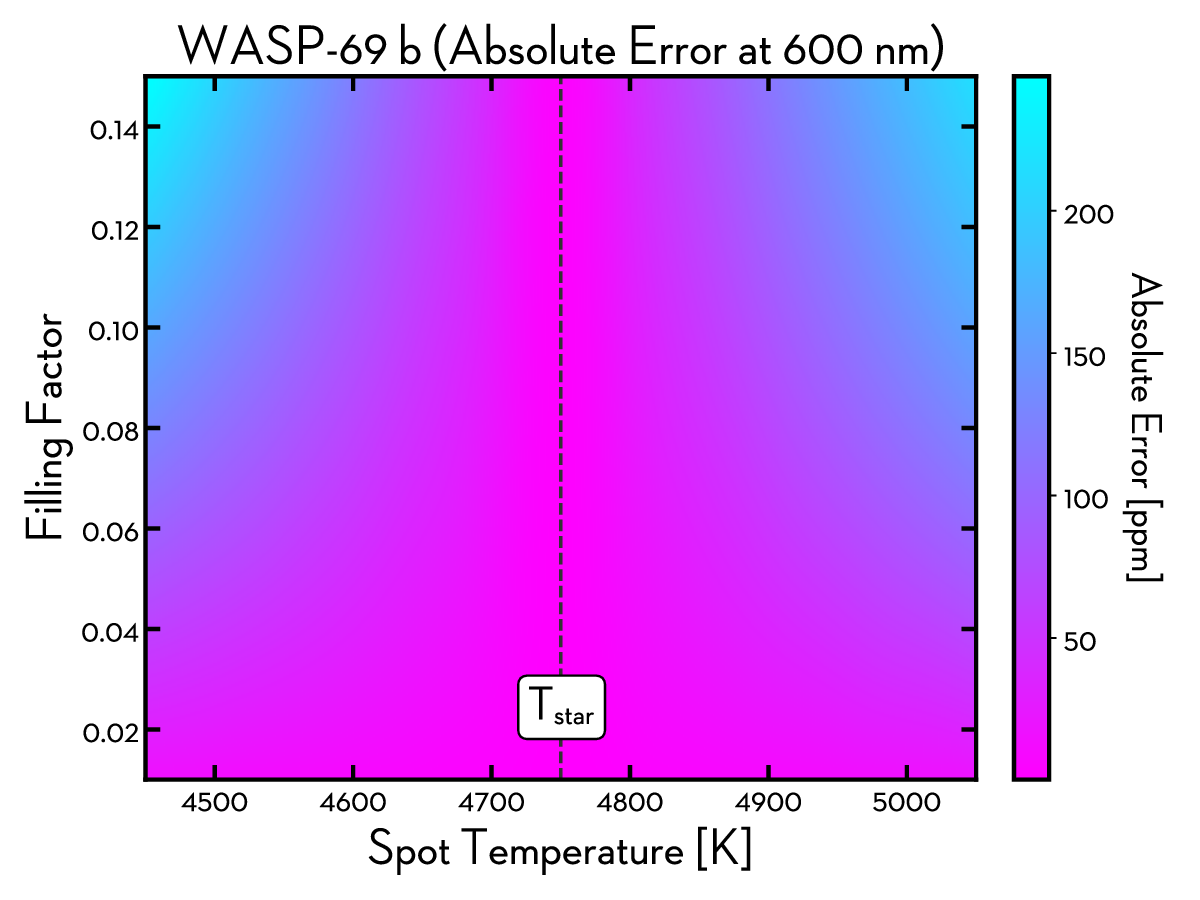}
    \caption{Two-dimensional parameter space exploration showing the discrepancies (\autoref{eq:err}) at a representative wavelength of 600\,nm. Each panel corresponds to a different exoplanetary system: LHS\,1140\,b (left panel), K2-18\,b (middle panel), and WASP-69\,b (right panel). Color scales indicate the difference between our model and the R-TLSE correction equation, as a function of filling factor and temperature contrast.}
    \label{fig:heatmaps}
\end{figure*}

Despite the fact that our simulations isolate stellar activity signals by fixing the temperature contrast or filling factor while varying the other, we further explore a scenario that systematically spans the full parameter space between these variables. This approach quantifies how degeneracies between temperature and spatial coverage influence model divergences. 
The results, illustrated in \autoref{fig:heatmaps} for all three exoplanets, reveal wavelength-dependent discrepancies between our method and the R-TLSE correction equation. For this analysis, we focus on 600\,nm, a wavelength selected to align with the lower limit of the JWST/NIRSpec observational range. 
According to this figure, we can clearly identify where the discrepancies are minimal (pink hues) and where they become more pronounced (blue). As expected, our model and the R-TLSE approximation tend to agree when the temperature of the star’s active regions is closer to its effective temperature. Conversely, the error increases toward the upper portion of the diagram, where filling factors are largest and the active regions exhibit the greatest temperature contrast relative to the stellar photosphere.

It is worth emphasising that in our simulations, the active region was placed as centrally as possible on the stellar disc without being occulted. In other words, the peak localised brightness at the disc centre was effectively suppressed by stellar activity. Nevertheless, our model also includes additional positional parameters that further complicate both the analysis and the potential mitigation of stellar contamination in transmission spectra. These include the number of starspots/faculae and their spatial location (closer to the centre or to the limb of the stellar disc). Each of these parameters can be varied independently, producing a vast array of combinations and potentially leading to significant computational demands when fitting to observational data. To illustrate this complexity, we focus exclusively on the exoplanet LHS\,1140\,b, varying the location of the active regions.

\begin{figure*}
    \centering
    \includegraphics[width=0.2\textwidth]{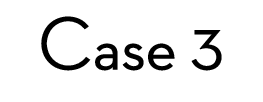}\\
    \includegraphics[width=0.49\textwidth]{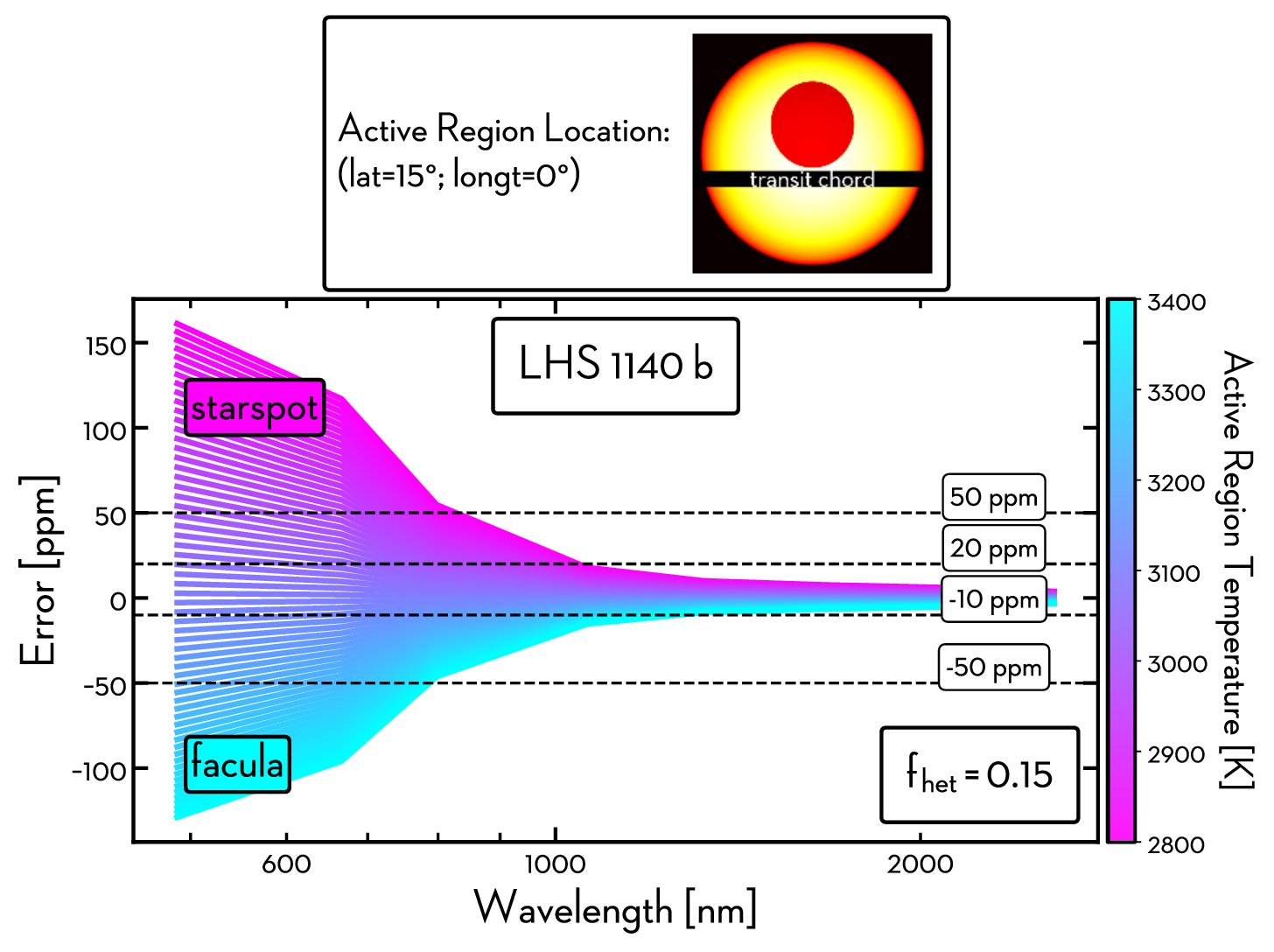}
    \includegraphics[width=0.49\textwidth]{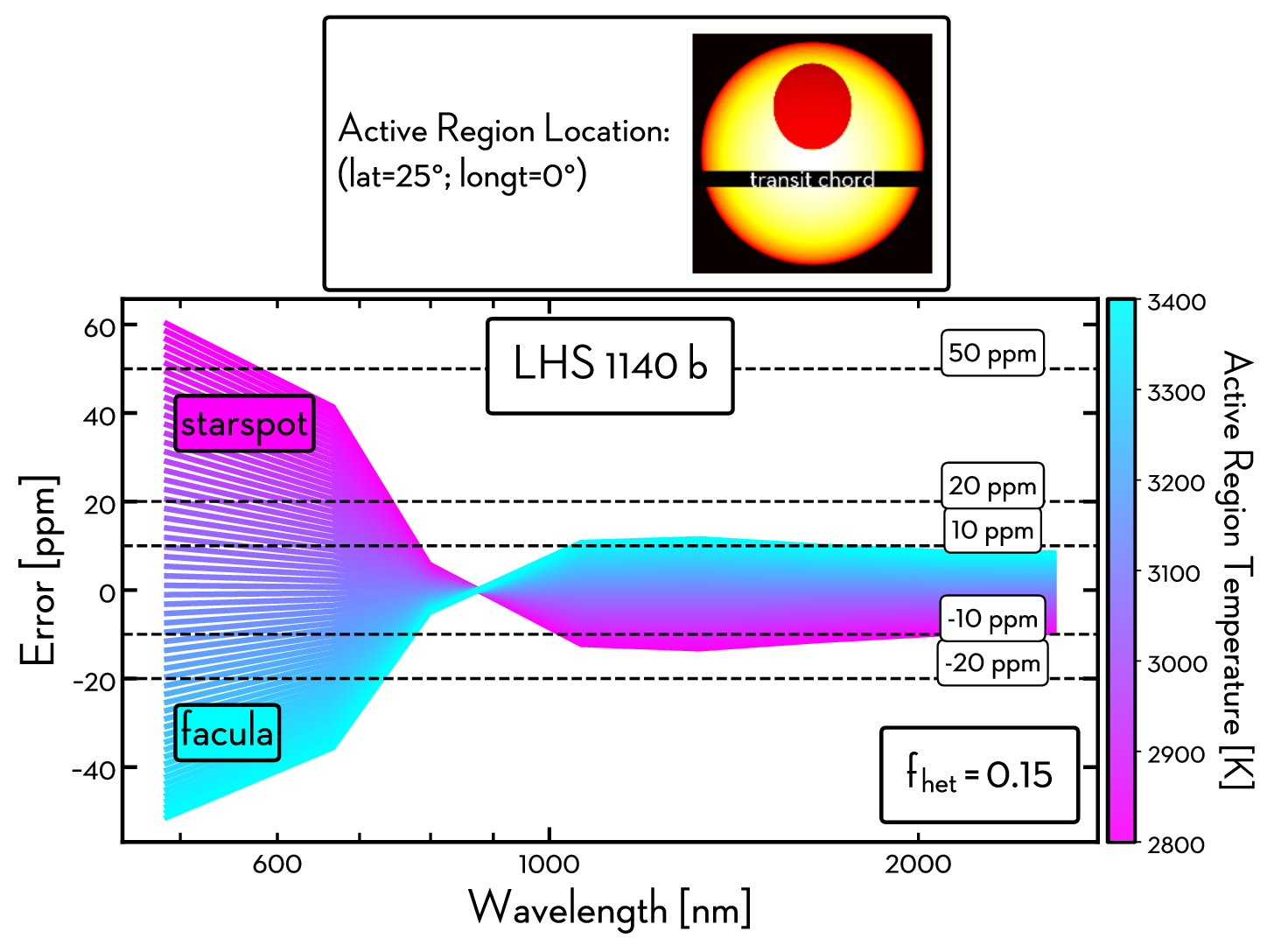}\\
    \includegraphics[width=0.49\textwidth]{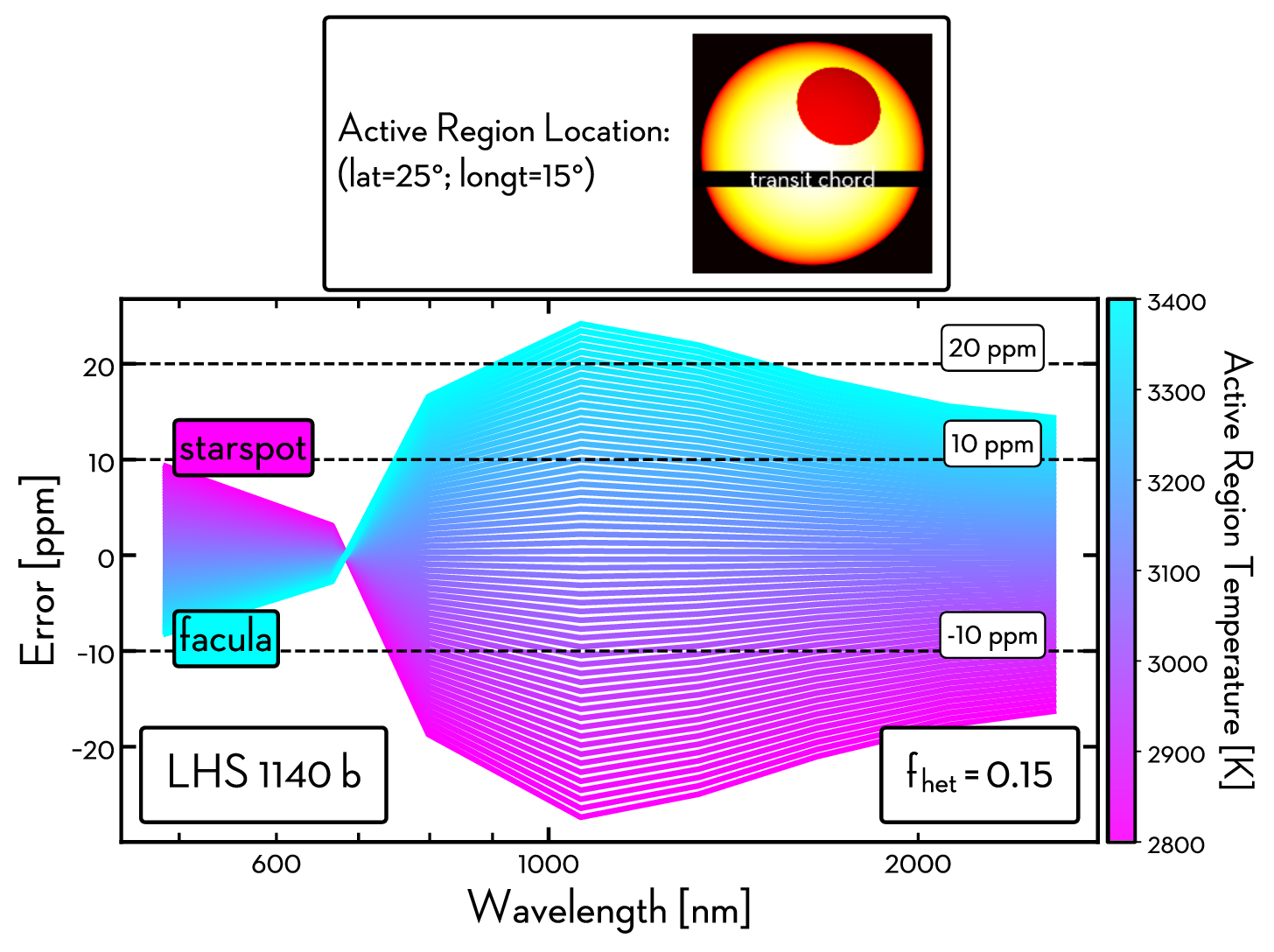}
    \includegraphics[width=0.49\textwidth]{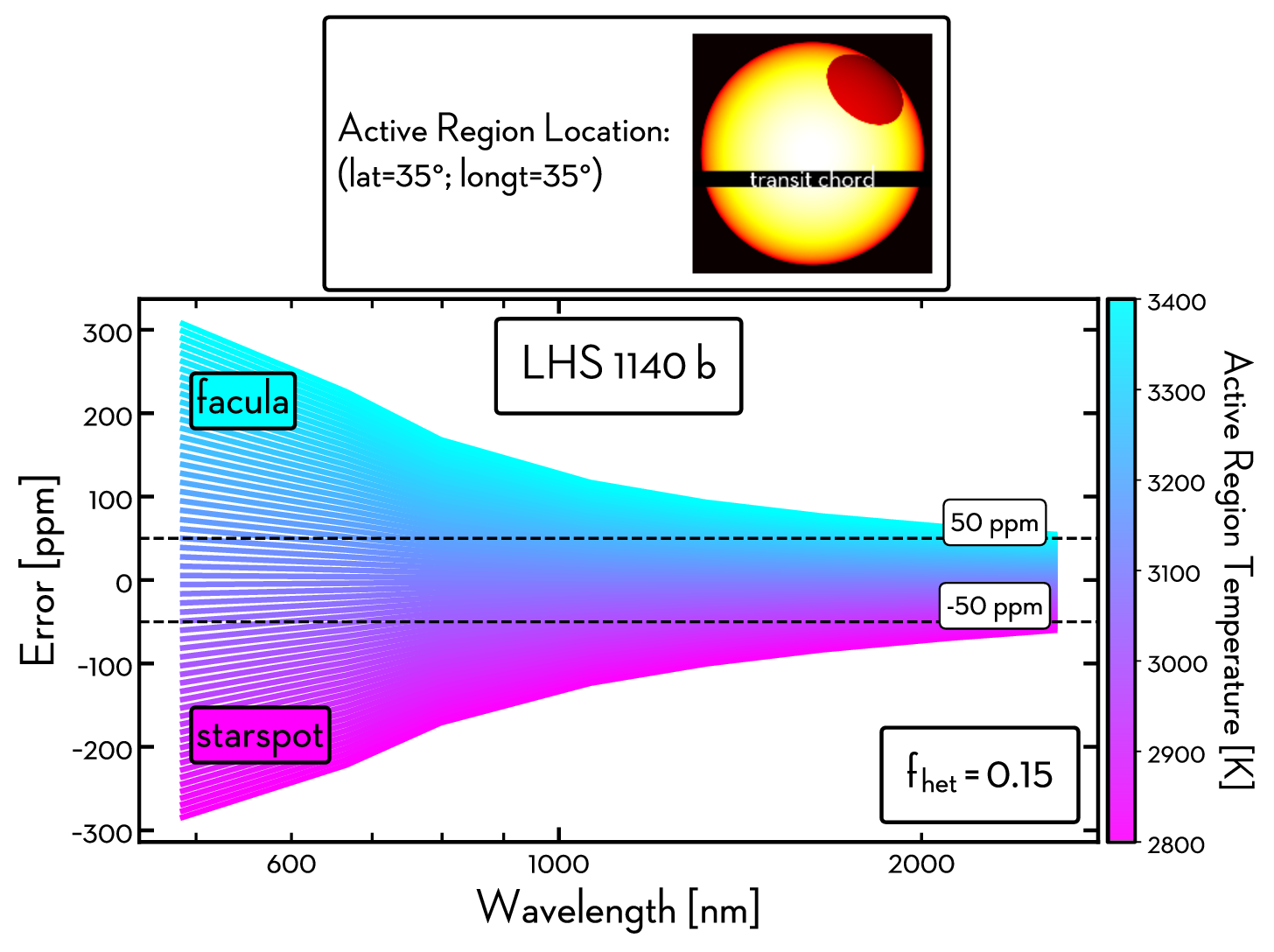}
    \caption{Effect of the active region's spatial position on the stellar disc for LHS\,1140\,b, illustrating the wavelength-dependent error between \texttt{ECLIPSE-X$\lambda$} and R-TLSE. Each curve corresponds to a different temperature of the active region (spot or facula). The four panels examines positions gradually shifting from close to disc centre (15$^\circ$ latitude, 0$^\circ$ longitude) toward the limb (35$^\circ$ latitude, 35$^\circ$ longitude). A sign inversion emerges at intermediate positions beyond the optical wavelengths (right panel, first row). The lower panels depict more extreme cases with active regions placed even closer to the limb, showing that the sign-inversion boundary moves to shorter wavelengths, significantly altering and reducing errors in the optical regime, while further limb positioning exacerbates R-TLSE inaccuracies due to enhanced limb darkening and foreshortening.
    Positive errors mean that R--TLSE predicts a shallower transit than ECLIPSE-X$\lambda$ (overcorrecting contamination), whereas negative errors mean that R--TLSE predicts a deeper transit (undercorrecting contamination).}
    \label{fig:LHS1140b_spot_locations}
\end{figure*}

In \autoref{fig:LHS1140b_spot_locations}, which corresponds to Case~3, we show the impact of the active region’s position on the stellar disc by systematically varying its location. 
To this end, we consider only the simulations with temperature variations, 
which are represented by the colours of faculae and starspots.
We start with the active region at 15$^\circ$ latitude and 0$^\circ$ longitude, gradually shifting to 35$^\circ$ latitude and 35$^\circ$ longitude, effectively moving it from the centre toward the stellar limb. At each step, we observe substantial discrepancies between our model and R-TLSE. 
Initially, the result is the same as in \autoref{fig:LHS1140b_err_lambda_ff_0p15.eps}, since the active region is placed exactly as in the previously discussed scenario. 
However, in the right panel of the upper row, where the latitude is increased by just 10$^\circ$, an intriguing effect emerges: the error associated with both starspots and faculae undergoes a sign reversal beyond the optical regime. As a result, in the NIR range, the R-TLSE approximation -- previously overestimating these contributions -- now underestimates the impact of stellar heterogeneities.

In the lower panels of \autoref{fig:LHS1140b_spot_locations}, we push our analysis to more extreme geometries by placing active regions closer to the stellar limb. In the left panel, the active region is positioned at a latitude of 25$^\circ$ and a longitude of 15$^\circ$, shifting the sign-inversion boundary toward shorter wavelengths and thereby substantially reducing the errors in the optical range (where the largest discrepancies were previously observed).
In the right panel of the second row, we consider an even more extreme configuration, with the active region located near both high latitude and high longitude -- practically on the stellar limb. Here, the sign inversion becomes fully established, the errors again escalate due to the pronounced influence of limb darkening and foreshortening. As the spot or facula moves to the limb, its projected area -- and thus its impact on the stellar flux -- becomes disproportionately large, significantly exacerbating the shortcomings of the R-TLSE-based equation.

Overall, these analyses, depicted in the panels of \autoref{fig:LHS1140b_spot_locations},  underscore that inaccuracies associated with the R-TLSE correction equation grow significantly when active regions are located near the centre or the stellar limb, due to the intricate interplay of limb darkening and stellar surface geometry. This highlights the necessity for self-consistent modelling of stellar contamination when interpreting transmission spectra, especially for exoplanets transiting active stars.

\subsection{Application to observational data: LHS\,1140\,b}
\label{sec:observational_LHS1140b}

\begin{figure*}
    \centering
    \includegraphics[width=0.99\textwidth]{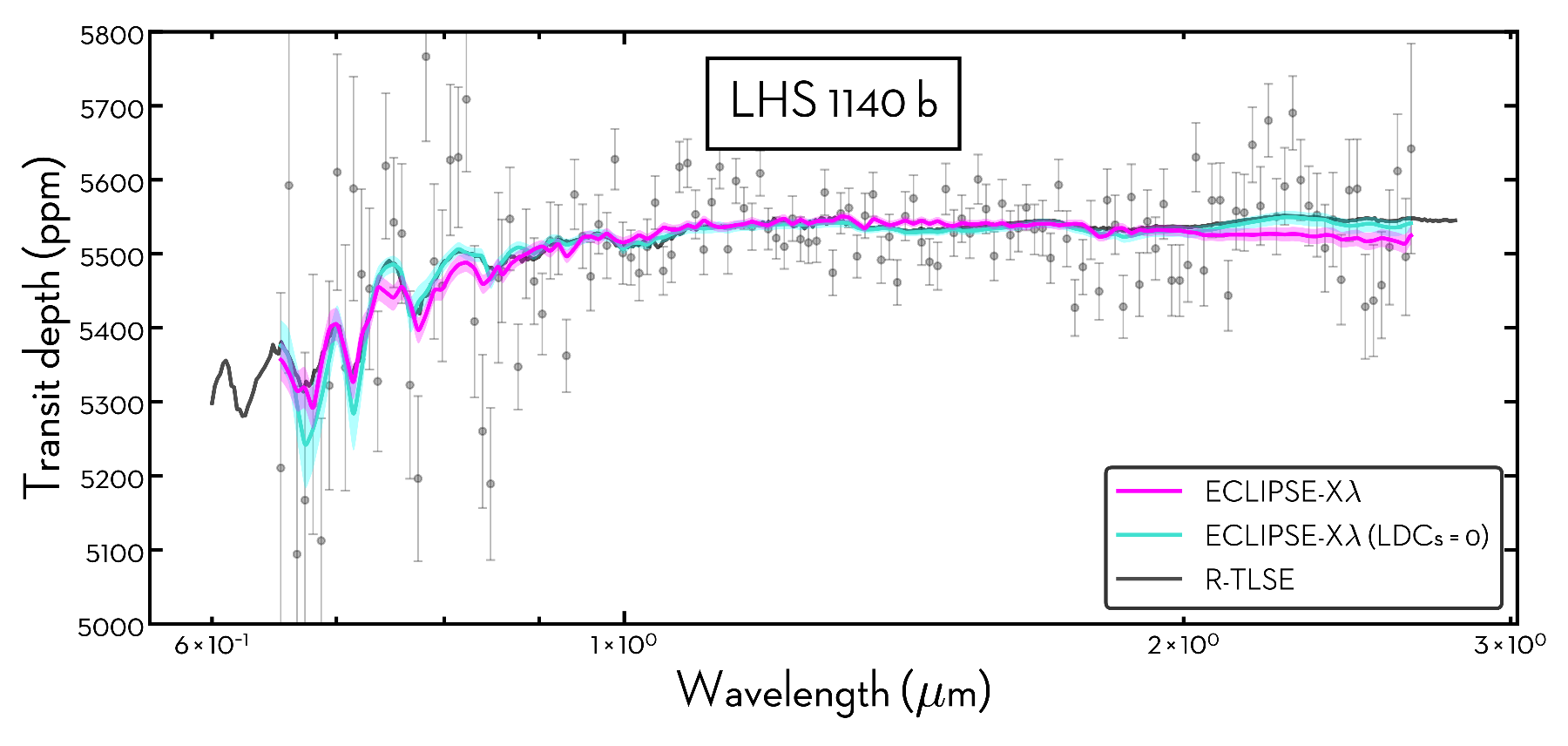}\\
    \includegraphics[width=0.9\textwidth]{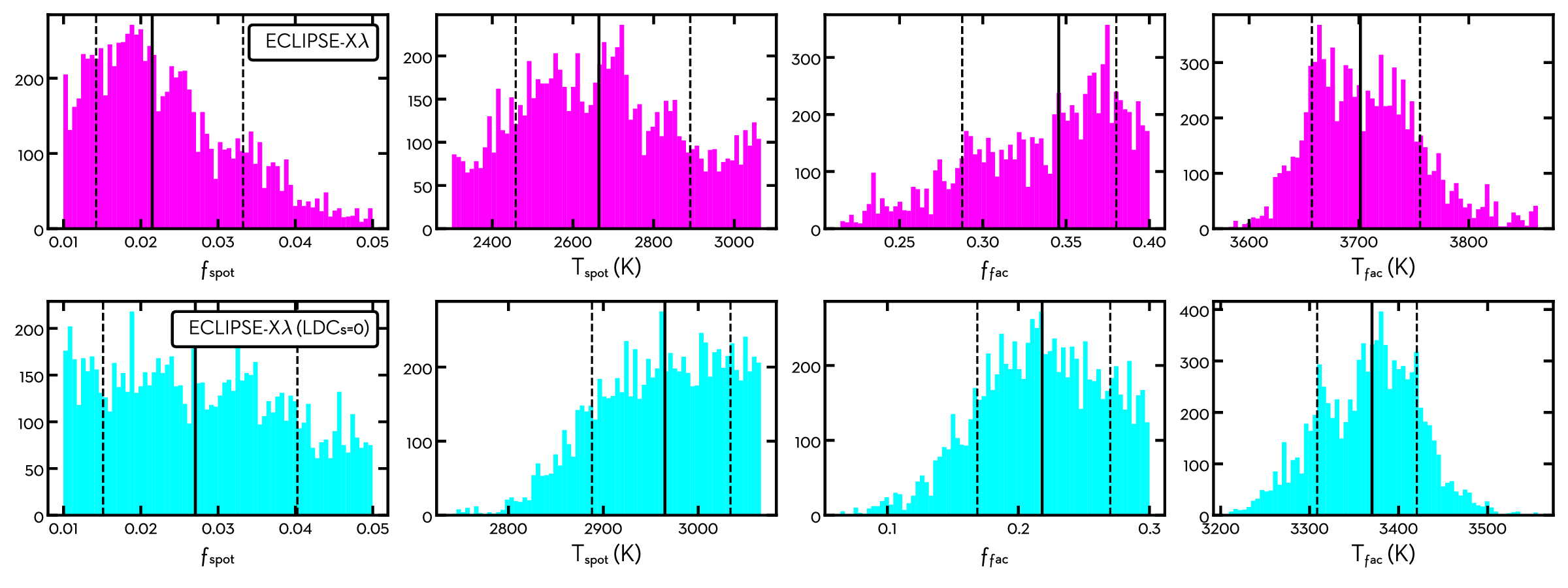}\\
    \caption{Comparison between \texttt{ECLIPSE-X$\lambda$} and the R--TLSE approximation for the JWST/NIRISS SOSS transmission spectrum of LHS\,1140\,b. \emph{Top:} Combined $R\!\sim\!100$ NIRISS spectrum (grey points with error bars), together with the best-fitting stellar–contamination models. The magenta curve shows the self-consistent \texttt{ECLIPSE-X$\lambda$} fit including wavelength-dependent limb darkening, the cyan curve shows the corresponding fit with limb darkening set to zero (LDCs\,=\,0), and the black curve reproduces the TLS-only stellar–contamination model from the R--TLSE reference retrieval \citep[see][]{Cadieux.et.al.2024ApJ...970L...2C}. 
    The small-scale structure along the model curves follows atomic and molecular absorption bands in the underlying PHOENIX spectra: TiO/VO and metal lines dominate in the optical, while broad features in the near-infrared are mainly shaped by H$_2$O bands and CO absorption complexes longwards of $\sim$2.2~$\mu$m, i.e. precisely the wavelength ranges where exoplanet transmission spectra are usually interpreted in terms of H$_2$O, CH$_4$, and CO absorption.
    \emph{Bottom:} One-dimensional marginal distributions for the stellar–contamination parameters in the ECLIPSE-X$\lambda$ fits, for the cases with limb darkening (top row, magenta) and with LDCs$=0$ (bottom row, cyan). From left to right, panels show the filling factor and temperature of spots ($f_{\mathrm{spot}}$, $T_{\mathrm{spot}}$) and faculae ($f_{\mathrm{fac}}$, $T_{\mathrm{fac}}$). Vertical lines indicate the median (solid) and 16th–84th percentile (dashed) intervals reported in Table~\ref{tab:stellar_contam}.}
    \label{fig:LHS1140b_ECLIPSE_TLSE-R}
\end{figure*}

The transmission spectrum analysed in this section is the combined JWST/NIRISS SOSS data set of LHS\,1140\,b, consisting of two transits reduced and binned to $R\!\sim\!100$ \citep{Cadieux.et.al.2024ApJ...970L...2C}\footnote{The observational spectrum was provided by the authors of \cite{Cadieux.et.al.2024ApJ...970L...2C}.} For reference, the R--TLSE column in \autoref{tab:stellar_contam} reproduces the TLS-only stellar-contamination solution reported by \cite{Cadieux.et.al.2024ApJ...970L...2C} at this resolution. We then fit the same $R\!\sim\!100$ spectrum with \texttt{ECLIPSE-X$\lambda$} by exploring a regular grid in the temperatures and filling factors of spots and faculae. For each grid point, \texttt{ECLIPSE-X$\lambda$} computes the corresponding disc-integrated photosphere + activity spectrum, folds it through the NIRISS/SOSS throughput and spectral response to match the $R\!\sim\!100$ channel grid, incorporates the transit light source (TLS) effect in the model transit depths, and evaluates the resulting transmission spectrum against the observations via a $\chi^{2}$ statistic. The best-fitting solutions for models with our nominal limb-darkening treatment and with limb darkening set to zero (LDCs\,=\,0) are shown in \autoref{fig:LHS1140b_ECLIPSE_TLSE-R} and summarised in \autoref{tab:stellar_contam}.

\begin{table*}
\renewcommand{\arraystretch}{1.3} 
\caption{Stellar–contamination retrieval summary for LHS\,1140\,b.}
\label{tab:stellar_contam}
\centering
\begin{tabular}{l c c c}
\hline\hline
Parameter &
R--TLSE$^{(a)}$ &
ECLIPSE-X$\lambda$ &
ECLIPSE-X$\lambda$ (LDCs=0) \\
\hline
\multicolumn{4}{c}{Stellar contamination}\\
\hline
$T_{\mathrm{spot}}$ (K) &
$3000^{+70}_{-70}$ &
$2660^{+230}_{-210}$ &
$2960^{+80}_{-70}$ \\
$T_{\mathrm{phot}}$ (K) &
$3070^{+40}_{-50}$ &
$3100^{+40}_{-50}$ &
$3100^{+40}_{-40}$ \\
$T_{\mathrm{fac}}$ (K) &
$3150^{+70}_{-90}$ &
$3700^{+50}_{-40}$ &
$3370^{+50}_{-60}$ \\
$f_{\mathrm{spot}}$ &
$0.04^{+0.02}_{-0.01}$ &
$0.021^{+0.012}_{-0.07}$ &
$0.027^{+0.013}_{-0.012}$ \\
$f_{\mathrm{fac}}$ &
$0.21^{+0.14}_{-0.12}$ &
$0.35^{+0.03}_{-0.06}$ &
$0.22^{+0.05}_{-0.05}$ \\
\hline
\multicolumn{4}{c}{Model statistics}\\
\hline
$\chi^{2}$ (d.o.f.) &
$235\,(136)$ &
$241\,(136)$ &
$237\,(136)$ \\
$\chi^{2}_{\mathrm{red}}$ &
$1.73$ &
$1.77$ &
$1.74$ \\
\hline
\end{tabular}
\tablefoot{
Columns compare the reference TLS-only stellar-contamination solution
(R--TLSE) with the ECLIPSE-X$\lambda$ fits including wavelength-dependent
limb darkening and with limb darkening set to zero (LDCs = 0).
The upper block lists the inferred temperatures and filling factors of
spots and faculae; the lower block reports the corresponding
goodness-of-fit statistics.
(a) The R--TLSE column reproduces the TLS-only stellar-contamination
results from the reference retrieval of \cite{Cadieux.et.al.2024ApJ...970L...2C}.
}
\end{table*}

As a first step, we compare the \texttt{ECLIPSE-X$\lambda$} retrieval with limb darkening switched off (LDCs\,=\,0) to the reference R--TLSE solution. When all limb-darkening coefficients are set to zero, the stellar disc becomes uniformly bright and the transit chord no longer samples any centre-to-limb intensity gradient. 
In this case, the pixelised \texttt{ECLIPSE-X$\lambda$} simulations approach the same disc-averaged description of stellar contamination that underlies the R--TLSE formalism. This modelling configuration is closely related to the situation discussed by \cite{Alexoudi.et.al.2020A&A...640A.134A}, who showed that removing the host-star limb darkening causes the impact-parameter degeneracy to disappear: the inferred planet-to-star radius ratio $R_{\mathrm{p}}/R_{\mathrm{s}}$ no longer exhibits any wavelength-dependent offset and remains constant across the spectrum. Adopting this limb-darkening-free configuration in \texttt{ECLIPSE-X$\lambda$} therefore eliminates that particular source of bias and isolates any residual differences between our model and R--TLSE.

\autoref{tab:stellar_contam} shows that, under these conditions, the stellar-contamination parameters recovered by \texttt{ECLIPSE-X$\lambda$} (LDCs\,=\,0) remain very close to the R--TLSE solution. The inferred photospheric temperatures agree at the $\sim 1\sigma$ level ($T_{\mathrm{phot}} = 3073^{+45}_{-54}$\,K for R--TLSE versus $3100^{+40}_{-40}$\,K for ECLIPSE-X$\lambda$ with LDCs$=0$), and the facular filling factors are essentially identical ($f_{\mathrm{fac}} = 0.21^{+0.14}_{-0.12}$ compared to $0.22^{+0.052}_{-0.050}$). A small but noticeable difference arises in the facular temperature, with \texttt{ECLIPSE-X$\lambda$} (LDCs\,=\,0) favouring hotter faculae ($T_{\mathrm{fac}} = 3370^{+50}_{-62}$\,K) than the R--TLSE retrieval ($3155^{+69}_{-91}$\,K), while keeping $f_{\mathrm{fac}}$ almost unchanged. A plausible interpretation is that this offset reflects geometric effects that are explicitly included in \texttt{ECLIPSE-X$\lambda$}, even when limb darkening is switched off: the model still accounts for foreshortening and for the localised position of active regions on the stellar disc, whereas the R--TLSE approximation assumes a spatially uniform, disc-averaged coverage. In our pixel-based framework, part of the geometric dilution of facular brightness can therefore be absorbed into a slightly higher intrinsic facular temperature rather than into a different filling factor, leading to the modest temperature shift seen in \autoref{tab:stellar_contam}, while preserving an almost identical overall level of stellar contamination.

When full limb darkening is included, the \texttt{ECLIPSE-X$\lambda$} fit departs more markedly from the R--TLSE solution unless the model is allowed to explore rather extreme facular coverages. As summarised in \autoref{tab:stellar_contam}, the R--TLSE analysis favours a relatively cool photosphere and modestly hotter faculae ($T_{\mathrm{phot}} = 3073^{+45}_{-54}$\,K and $T_{\mathrm{fac}} = 3155^{+69}_{-91}$\,K, with $f_{\mathrm{fac}} = 0.21^{+0.14}_{-0.12}$). By contrast, in the fully limb-darkened \texttt{ECLIPSE-X$\lambda$} simulations, where we allow unocculted faculae to occupy up to $f_{\mathrm{fac}} \leq 0.40$ of the visible hemisphere, the optimiser naturally drives the solution towards very hot and spatially extended faculae. the best-fitting self-consistent model for LHS\,1140\,b converges to a photospheric temperature very similar to the R--TLSE value ($T_{\mathrm{phot}} = 3099^{+44}_{-47}$\,K), but with much hotter faculae ($T_{\mathrm{fac}} = 3701^{+54}_{-44}$\,K, $\Delta T_{\mathrm{fac}} \approx 600$\,K) covering $f_{\mathrm{fac}} = 0.35^{+0.03}_{-0.06}$ of the stellar disc. This configuration lies close to the geometric limit at which active regions would start intersecting the transit chord, and it yields a reduced $\chi^{2}$ for the limb-darkened fit that is comparable to the R--TLSE solution, at the expense of invoking a very large population of unocculted faculae.

This behaviour is readily understood once we account for the impact of limb darkening on the baseline transit geometry. Even in the absence of stellar heterogeneities, a limb-darkened star does not, in general, produce a perfectly flat transmission spectrum: for a nearly central transit, the combination of impact parameter and centre-to-limb intensity variations naturally induces a modest positive slope in $R_{\mathrm{p}}/R_{\mathrm{s}}$ at optical wavelengths, as highlighted by \cite{Alexoudi.et.al.2020A&A...640A.134A}. In the R--TLSE (or LDCs\,=\,0) limit, this geometric effect is explicitly removed by construction, because the stellar disc is assumed to be uniformly bright, and the baseline spectrum is flat in the absence of active regions. Unocculted faculae then act on top of this flat reference and directly imprint a net slope. In the fully limb-darkened \texttt{ECLIPSE-X$\lambda$} framework, on the other hand, we already start from a positively sloped baseline set by the transit geometry; adding unocculted faculae must first compensate for this geometrically induced trend before it can reproduce any additional slope of opposite sign. The optimiser therefore tends to increase $T_{\mathrm{fac}}$ and to push $f_{\mathrm{fac}}$ towards the upper bound of the prior in order to counteract the limb-darkening-driven slope. However, foreshortening and the localised nature of the active regions limit how much extra bias they can introduce into the disc-integrated flux. As a result, R--TLSE-like slopes can be reproduced in the fully limb-darkened simulations only in the extreme regime of very hot faculae covering $\gtrsim 30$--40\% of the visible hemisphere. More moderate facular configurations naturally produce a shallower contamination signal, implying that any residual optical slope is likely to include a genuine atmospheric contribution, potentially combined with a less extreme level of stellar contamination, rather than being explained solely by unocculted faculae.

\section{Summary and conclusions}
\label{sec:conclusions}

This study demonstrates that the \cite{Rackham.et.al.2018ApJ...853..122R} Transit Light Source Effect (R--TLSE) approximation introduces systematic, wavelength-dependent biases in exoplanet transmission spectra, especially at optical wavelengths where limb-darkening gradients and transit geometry play a dominant role. Through idealised, noise-free simulations (Section~\ref{sec:idealised_model_comparison}), we compare the disc-averaged R--TLSE correction to our self-consistent \texttt{ECLIPSE-X$\lambda$} framework, which incorporates wavelength-dependent limb darkening, the spatial distribution of active regions, and the full transit geometry, and we quantify system-specific discrepancies for three archetypal planets. We then apply both approaches to the JWST/NIRISS SOSS observations of LHS~1140\,b (Section~\ref{sec:observational_LHS1140b}), using the data as a testbed for assessing the practical impact of these differences on atmospheric interpretation. Overall, our analysis shows that the scale of the R--TLSE-induced errors must be assessed on a case-by-case basis.

For LHS~1140\,b, a super-Earth orbiting an M dwarf, we find
peak discrepancies of 170\,ppm due to starspots and 140\,ppm due to faculae at 500\,nm, driven primarily by the R--TLSE's neglect of optical limb-darkening contrasts. These errors decrease to below 10\,ppm at near-infrared (NIR) wavelengths above 1000\,nm, where limb-darkening effects are weaker, rendering R--TLSE-based retrievals more reliable. However, in the optical regime (500–800\,nm), R--TLSE inaccuracies become largest and may contribute to marginal atmospheric detections. Notably,
the spectral region supporting the tentative N$_2$ signal reported by \cite{Cadieux.et.al.2024ApJ...970L...2C} coincides with wavelengths where R--TLSE inaccuracies are expected to peak and with the part of the spectrum that exhibits the largest observational uncertainties and dispersion, making any N$_2$ claim especially sensitive to the adopted stellar-contamination model.

In the case of K2-18\,b, a mini-Neptune also orbiting an M dwarf, the idealised simulations show that in the optical range (500–600\,nm) the R--TLSE approximation and the self-consistent model can differ by about 80–100\,ppm, comparable to the amplitude of key atmospheric features in sub-Neptune spectra. The planet’s mid-latitude transit geometry ($b\simeq0.65$) places the system in a ``flat-geometry'' regime in which a spotless, limb-darkened star already yields nearly wavelength-independent transit depths. In this configuration, modest chromatic stellar contamination or weak atmospheric gradients can produce shallow slopes that are comparable to current observational uncertainties, making it difficult to distinguish genuinely high-mean-molecular-weight, cloud-dominated atmospheres from more extended envelopes combined with residual stellar heterogeneity. At near-infrared wavelengths ($\lambda\gtrsim1.5\,\mu$m), however, \texttt{ECLIPSE-X$\lambda$} and R--TLSE agree to within $\sim$10\,ppm, consistent with previous analyses showing that methane detections based on R--TLSE-corrected spectra are robust, while inferences for secondary species such as CO$_2$ remain more vulnerable to stellar-contamination systematics.

The largest discrepancies are found for WASP-69\,b, a hot Jupiter orbiting a K-type star. Around 500\,nm the idealised simulations yield differences of up to $\sim$400\,ppm, and the errors remain above 50\,ppm up to roughly 1500\,nm, only dropping to the 10–50\,ppm level at longer wavelengths. This behaviour is amplified by the planet’s mid-latitude impact parameter ($b\approx0.66$–0.70) and by the large facular filling factors inferred from R--TLSE-based retrievals, which favour $f_{\mathrm{fac}}\sim30\%$ with $\Delta T_{\mathrm{fac}}\approx230$\,K in VLT/FORS2 analyses and even more extreme combinations such as $f_{\mathrm{spot}}=0.55$, $f_{\mathrm{fac}}=0.15$ in GTC/OSIRIS spectra. In our pixel-resolved framework, such large coherent active regions on the visible hemisphere would be difficult to reconcile with the mid-latitude transit chord without substantial occultations, strongly suggesting that the disc-averaged R--TLSE prescription is absorbing part of the atmospheric signal into the stellar-contamination term and thereby overestimating the facula filling factor. Together with independent evidence for high-altitude aerosols from JWST eclipse observations, this points to a scenario in which both genuine atmospheric opacity and moderate stellar heterogeneity shape the observed optical–NIR slopes.

The application to the JWST/NIRISS SOSS transmission spectrum of LHS~1140\,b provides a concrete observational test of these trends. When limb darkening is artificially set to zero (LDCs\,=\,0), \texttt{ECLIPSE-X$\lambda$} recovers stellar-contamination parameters that are fully consistent with the reference R--TLSE, TLS-only, solution, confirming that both approaches agree in the disc-averaged limit. Once wavelength-dependent limb darkening is restored, however, reproducing an R--TLSE-like optical slope requires very hot faculae ($\Delta T_{\mathrm{fac}}\simeq600$\,K) covering $f_{\mathrm{fac}}\simeq0.35$ of the visible hemisphere. For the transit configuration of LHS~1140\,b, this corresponds to an equivalent circular facular region with radius $R_{\mathrm{fac}}\simeq0.6\,R_\star$ on the stellar disc, already close to the limit where such regions would begin to intersect the transit chord. Facular filling factors of this order are already extreme even for active M dwarfs, especially if no part of the active regions is occulted during transit. If the true facular coverage is more modest, then stellar contamination alone would be unlikely to reproduce the full amplitude of any residual optical slope, and a contribution from the planetary atmosphere could naturally complement a smaller facular signal.

Accurate atmospheric retrievals, therefore, require self-consistent models that incorporate limb darkening, realistic spatial distributions of active regions, and an explicit treatment of the transit geometry. Multi-wavelength observations spanning both optical and NIR regimes are essential to break degeneracies between stellar and planetary signals: optical data are needed to constrain the chromatic imprint of stellar activity, while NIR measurements are crucial to anchor molecular bands where R--TLSE biases are smaller. Although JWST/NIRSpec extends into the red optical ($\sim$600\,nm), limb-darkening effects remain non-negligible at those wavelengths, so combining JWST spectroscopy with HST or high-precision ground-based optical data, together with dedicated stellar-activity monitoring, will be vital for robust spectral interpretation. Ultimately, stellar activity is not merely observational noise but a structured, time-variable and wavelength-dependent source of contamination that demands comprehensive modelling. Our results strongly support the routine use of advanced stellar-contamination frameworks, such as \texttt{ECLIPSE-X$\lambda$}, at least as a validation step for R--TLSE-based retrievals, particularly when interpreting subtle or marginal spectral features that are most vulnerable to unmodelled stellar biases.

\begin{acknowledgements}
VYDS and AV acknowledge the partial financial support received from Brazilian FAPESP grants \#2021/14897-9, \#2018/04055-8 \#2021/02120-0, and \#2024/03652-3, as well as CAPES and MackPesquisa funding agencies. This research was carried out at the Jet Propulsion Laboratory, California Institute of Technology, under a contract with the National Aeronautics and Space Administration (80NM0018D0004).
We thank the authors of \cite{Cadieux.et.al.2024ApJ...970L...2C} for kindly providing the observational data of LHS~1140\,b used in this work.
\end{acknowledgements}

%
\bibliographystyle{bibtex/aa.bst} 
\bibliography{references.bib} 

\begin{appendix}

\section{Supplementary Figures}
\label{sec:appendix}

In this appendix, we provide additional figures to complement and expand upon the analyses presented in the main text. The supplementary plots shown here systematically explore how the differences between the self-consistent model \texttt{ECLIPSE-X$\lambda$} and the TLSE approximation depend on key parameters, such as the filling factor and the temperature contrast, for the exoplanetary systems K2-18\,b and WASP-69\,b. These results highlight how the differences observed across systems orbiting different types of stars (M and K) can influence the systematic biases arising from simplified stellar-contamination treatments.

We also include posterior distributions for the stellar–contamination parameters of LHS~1140\,b, both for the fully limb-darkened fit and for the configuration with limb darkening set to zero (LDCs\,=\,0). These corner plots display the one- and two-dimensional marginals for the spot and facula filling factors and temperatures, as well as the photospheric temperature, with the corresponding numerical summaries reported in \autoref{tab:stellar_contam}.

\begin{figure*}
    \centering
    \includegraphics[width=0.99\textwidth]{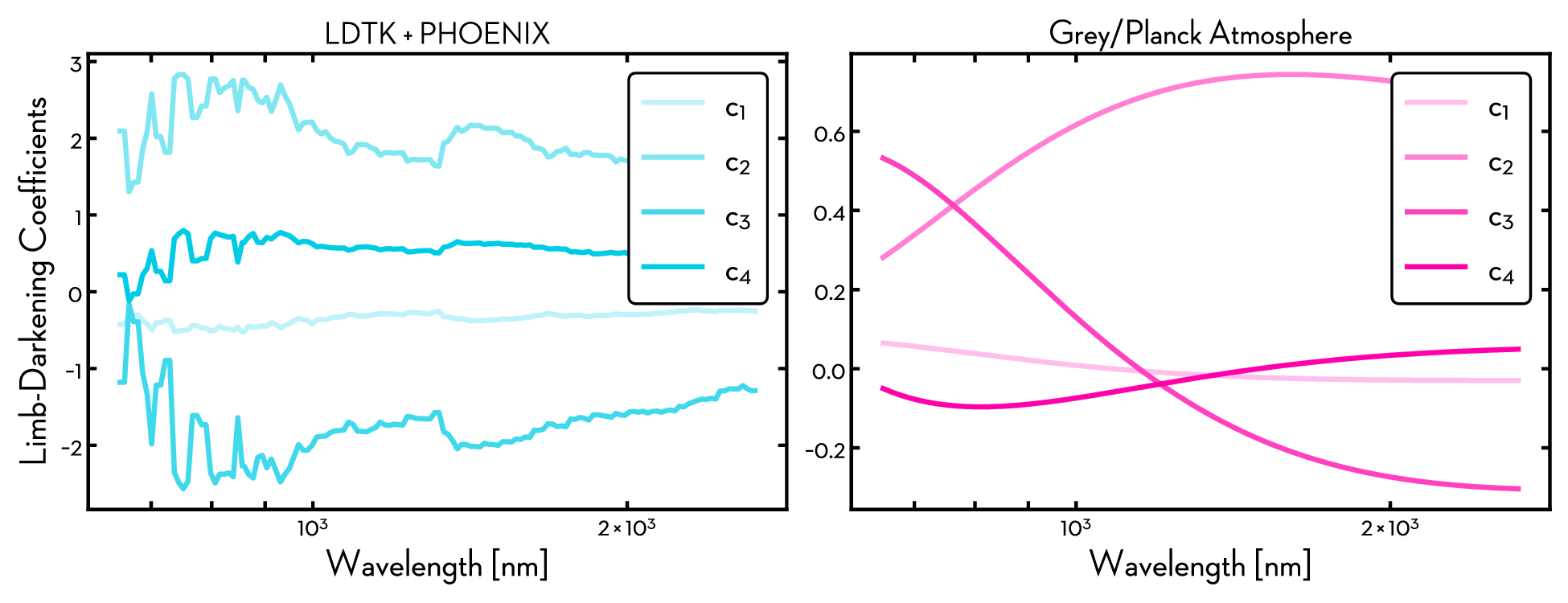}\\
    \caption{Wavelength-dependent limb-darkening coefficients for LHS\,1140\,b. \emph{Left:} Coefficients $c_{1}$–$c_{4}$ of the four-parameter law obtained with LDTK \citep{Parviainen.and.Aigrain.2015MNRAS.453.3821P} using PHOENIX specific intensities, showing strong small-scale structure inherited from spectral lines. \emph{Right:} Corresponding coefficients from our grey/Planck atmosphere approximation, yielding a smooth, line-free wavelength dependence that we adopt for the self-consistent ECLIPSE-X$\lambda$ simulations.}
    \label{fig:LDCs}
\end{figure*}

\begin{figure*}
    \centering
    \includegraphics[width=0.49\textwidth]{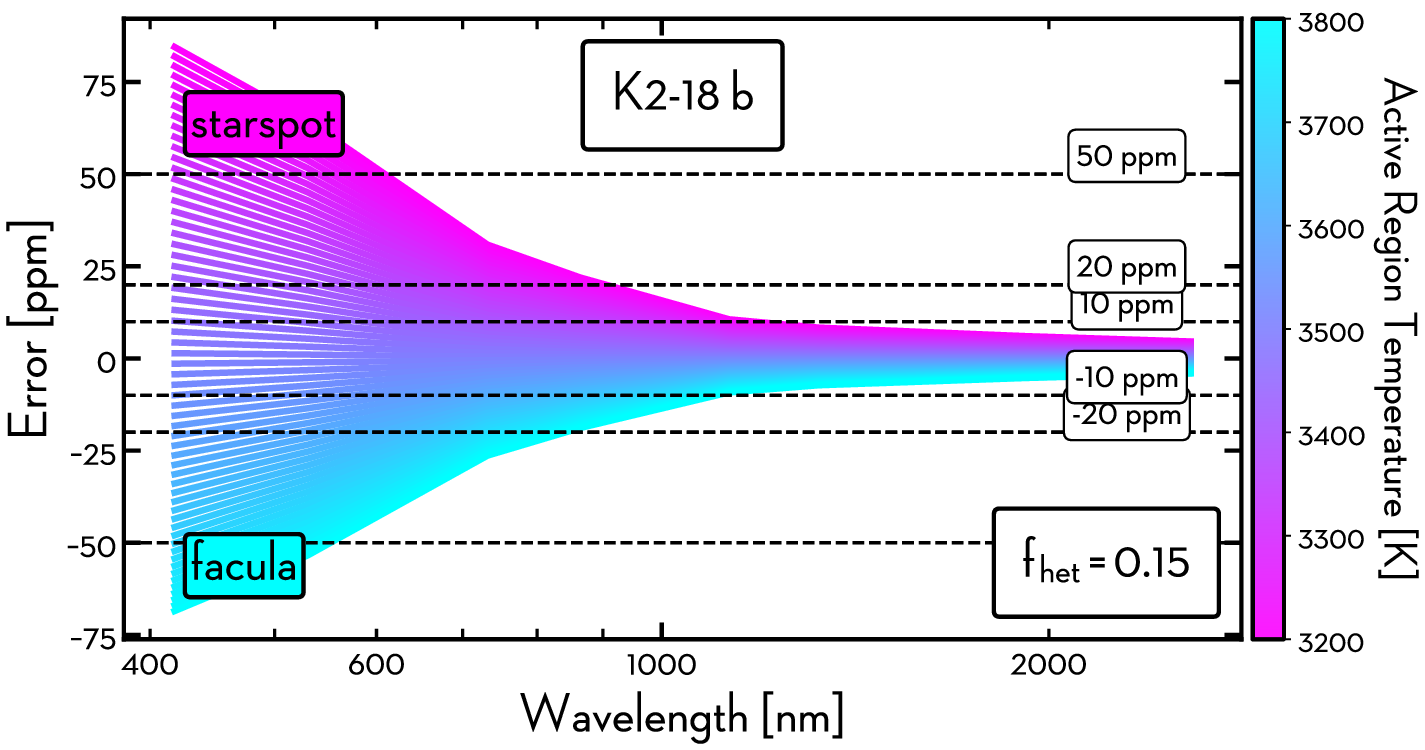}
    \includegraphics[width=0.49\textwidth]{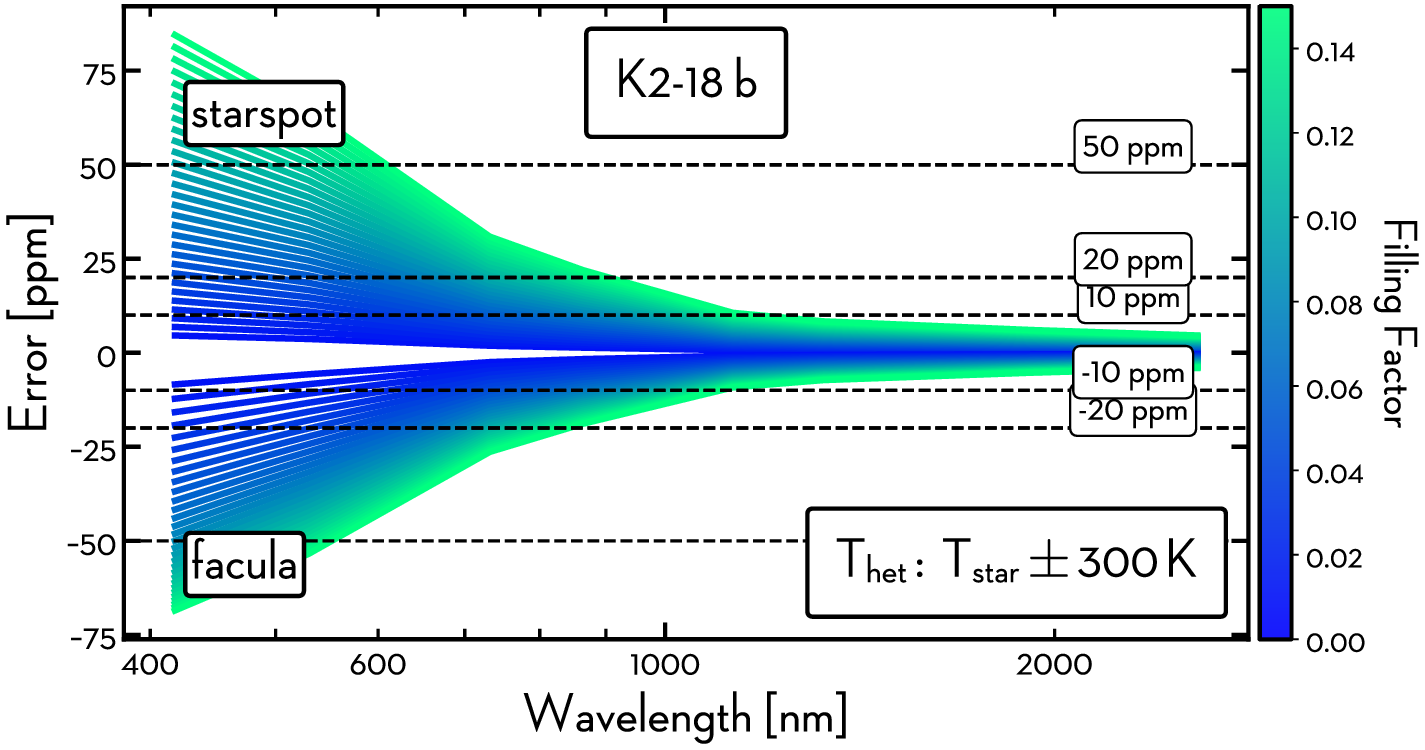}\\
    \includegraphics[width=0.49\textwidth]{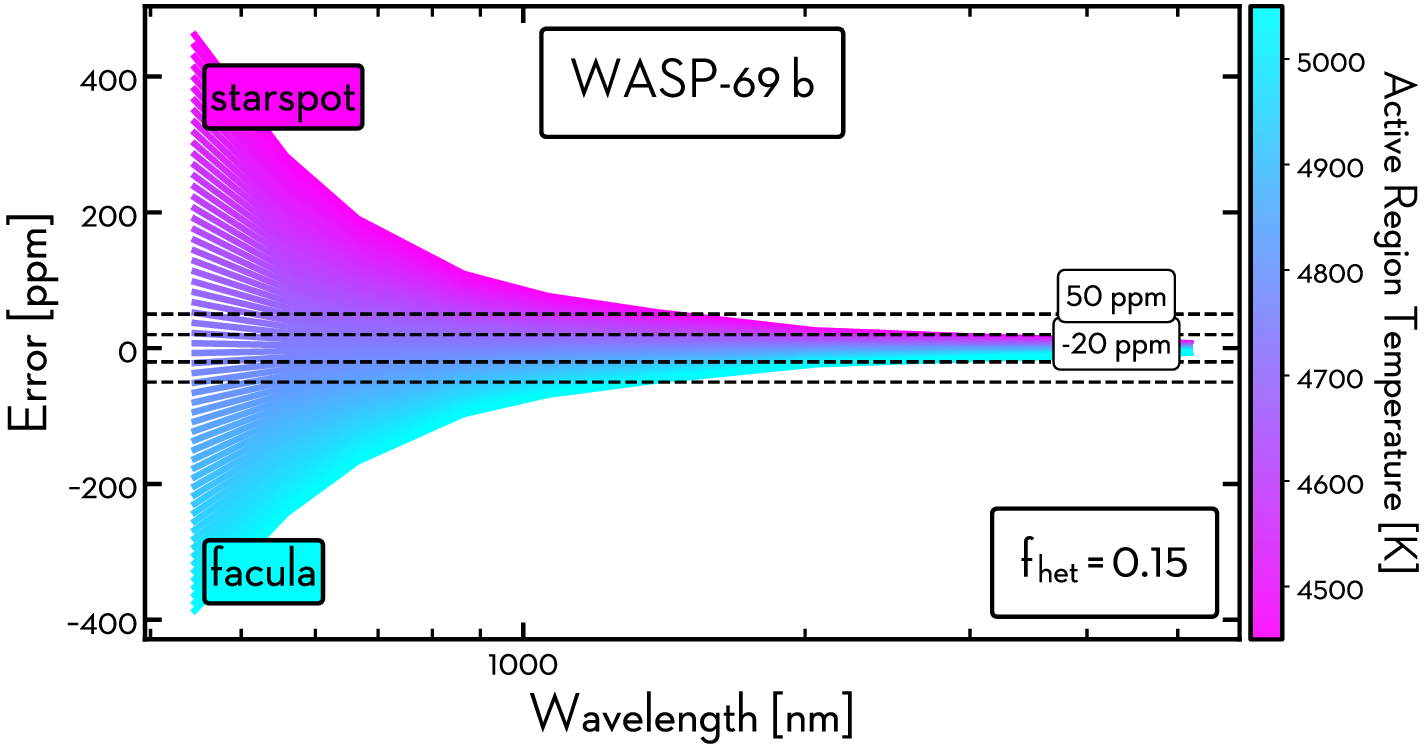}
    \includegraphics[width=0.49\textwidth]{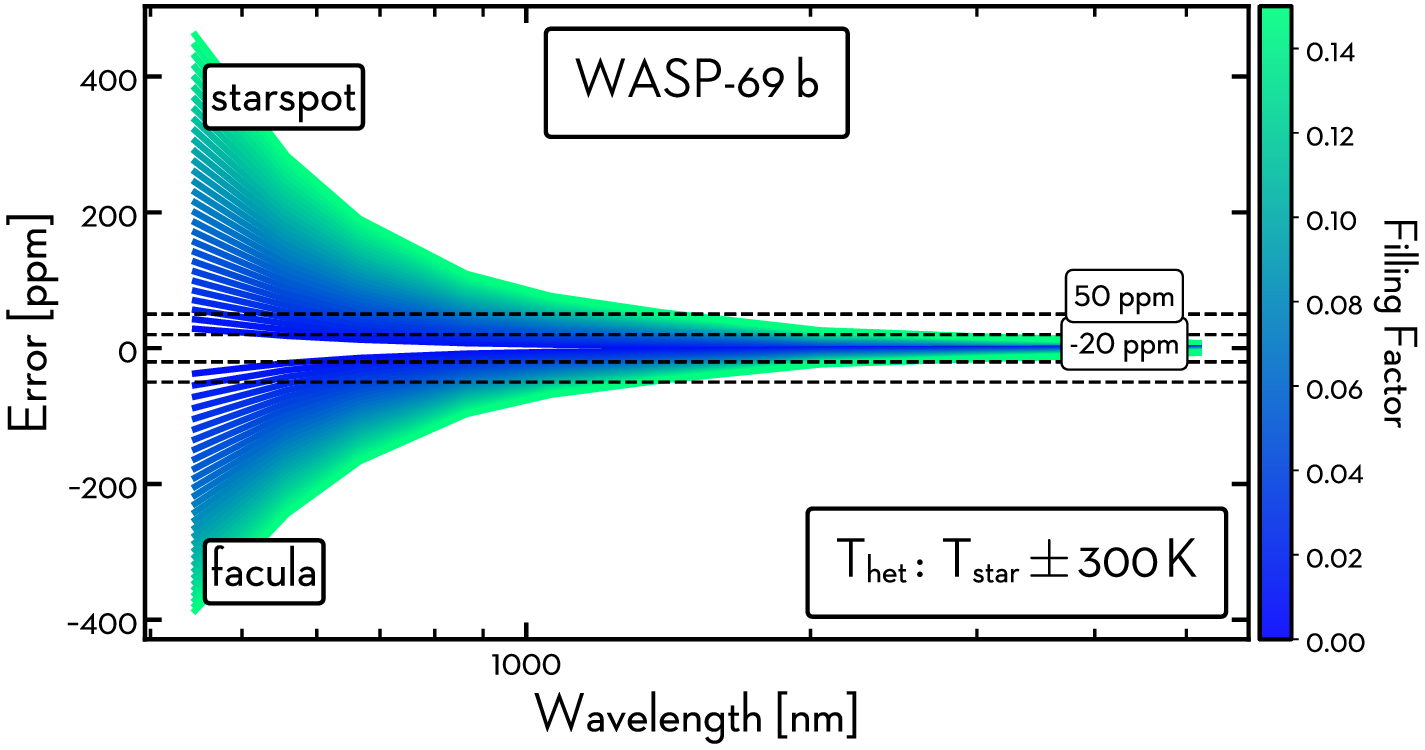}
    \caption{Supplementary analysis analogous to Figures \ref{fig:LHS1140b_err_lambda_ff_0p15.eps} and \ref{fig:LHS1140b_err_lambda_T_300}, but for the exoplanets K2-18\,b (upper panels) and WASP-69\,b (lower panels). The panels systematically explore the effect of varying the temperature contrast (left panels) and the filling factor (right panels), respectively. These supplementary figures confirm trends identified for LHS\,1140\,b, while also highlighting the distinct stellar host type dependency (M-dwarf vs. K-type stars) of systematic biases between \texttt{ECLIPSE-X$\lambda$} and TLSE. The dashed lines serve as reference markers, providing a visual support for interpreting variations in the data.}
    \phantomsection
    \label{fig:K2-18b_WASP-17b_err_lambda}
\end{figure*}

\begin{figure*}
    \centering
    \includegraphics[width=0.99\textwidth]{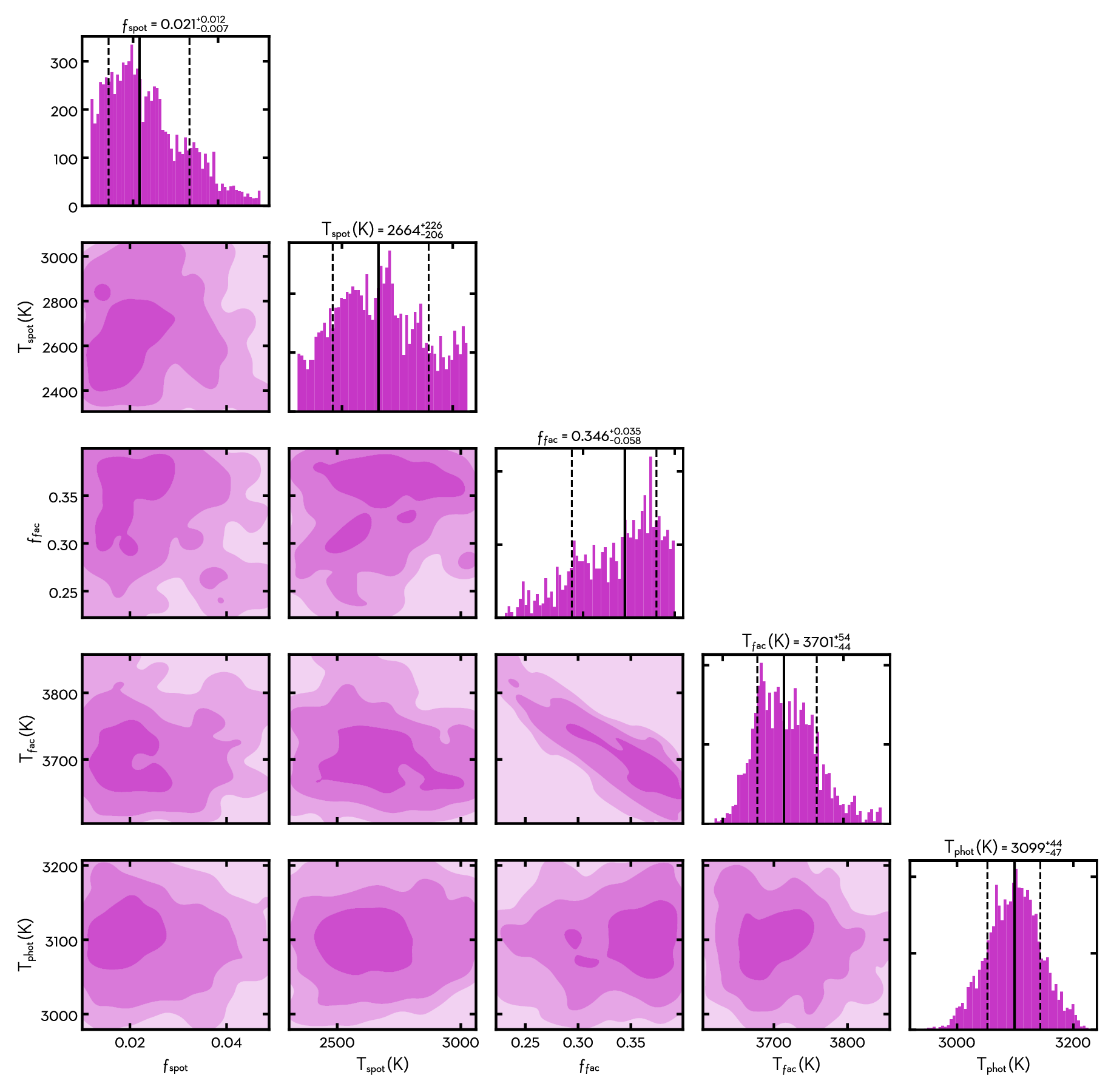}
    \caption{Posterior distributions for the stellar–contamination parameters of LHS~1140\,b from the \texttt{ECLIPSE-X$\lambda$} fit with limb darkening included. The panels show the one- and two-dimensional marginals for the spot filling factor $f_{\mathrm{spot}}$, spot temperature $T_{\mathrm{spot}}$, facula filling factor $f_{\mathrm{fac}}$, facula temperature $T_{\mathrm{fac}}$, and photospheric temperature $T_{\mathrm{phot}}$. Diagonal panels display the 1D histograms with the median and 16th–84th percentile intervals marked by solid and dashed vertical lines, respectively. Off-diagonal panels show the corresponding 2D credible regions, with darker shades indicating higher posterior density. Numerical summaries of these posteriors are reported in \autoref{tab:stellar_contam}.}
    \label{fig:}
\end{figure*}

\begin{figure*}
    \centering
    \includegraphics[width=0.99\textwidth]{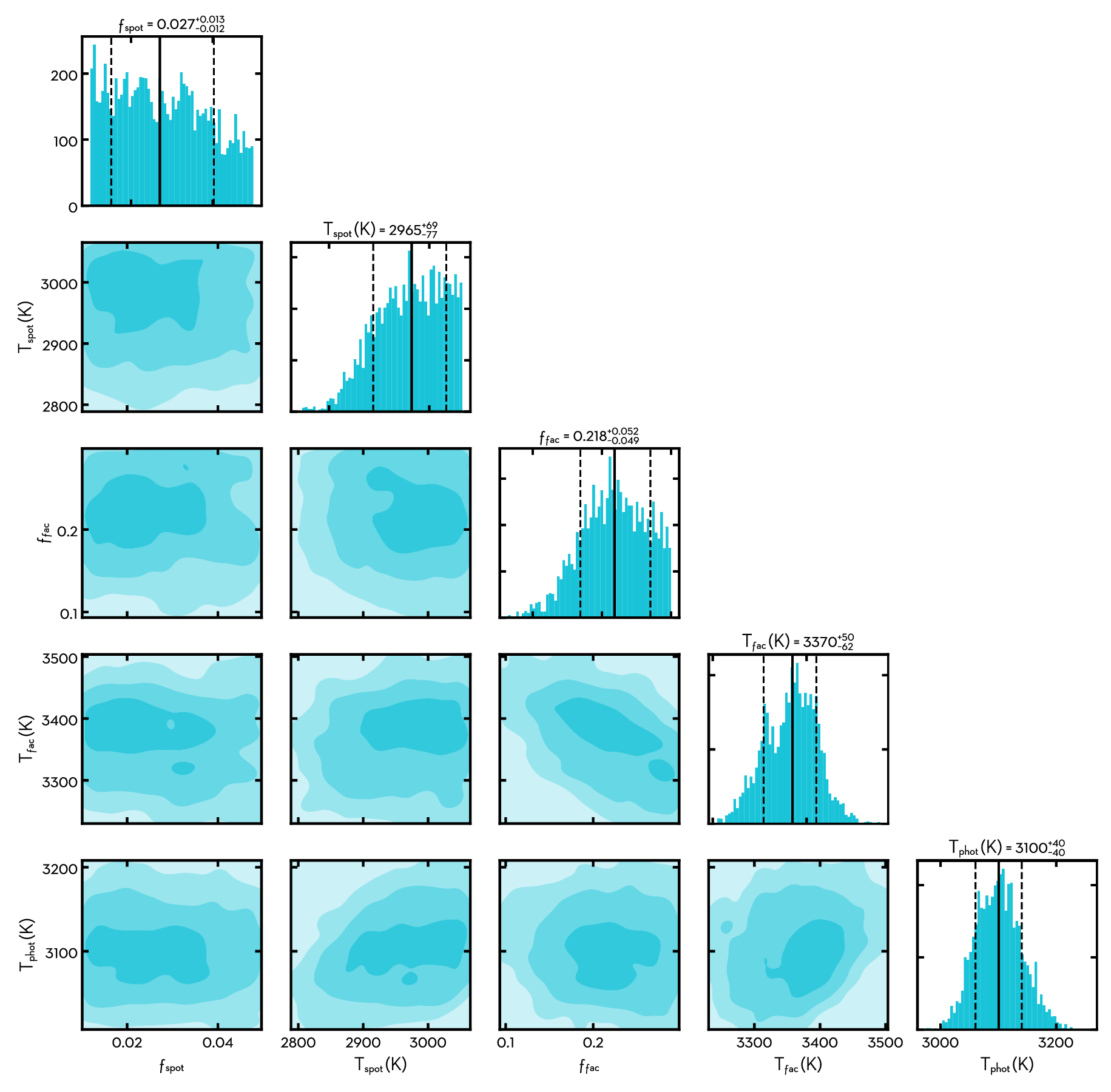}
    \caption{Posterior distributions for the stellar–contamination parameters of LHS~1140\,b from the \texttt{ECLIPSE-X$\lambda$} fit with limb darkening set to zero (LDCs\,=\,0). The panels show the one- and two-dimensional marginals for the spot filling factor $f_{\mathrm{spot}}$, spot temperature $T_{\mathrm{spot}}$, facula filling factor $f_{\mathrm{fac}}$, facula temperature $T_{\mathrm{fac}}$, and photospheric temperature $T_{\mathrm{phot}}$. Diagonal panels display the 1D histograms with the median and 16th–84th percentile intervals marked by solid and dashed vertical lines, respectively. Off-diagonal panels show the corresponding 2D credible regions, with darker shades indicating higher posterior density. Numerical summaries of these posteriors are reported in \autoref{tab:stellar_contam}.}
    \label{fig:LHS1140b_corner_LDCs_0}
\end{figure*}

\end{appendix}
\end{document}